\documentclass[a4paper,11pt]{article}
\usepackage{jheppub}

\usepackage[utf8]{inputenc}
\usepackage{graphicx,cancel,float}
\usepackage{amssymb}
\usepackage{textcomp}
\usepackage{amsmath}
\usepackage{tabularx}
\usepackage{bm}
\usepackage{times}
\usepackage{color}
\usepackage{slashed}
\usepackage{multirow}
\usepackage{verbatim}
\usepackage{epsfig,epstopdf}
\usepackage{subfigure}
\allowdisplaybreaks[4]

\def\lsim{\mathrel{\raise.3ex\hbox{$<$\kern-.75em\lower1ex\hbox{$\sim$}}}}
\def\gsim{\mathrel{\raise.3ex\hbox{$>$\kern-.75em\lower1ex\hbox{$\sim$}}}}

\definecolor{orange}{rgb}{1,0.5,0}

\newcommand{\minigraph}[5][0.25in]{\begin{minipage}{#2}\begin{center}\includegraphics[width=#2]{#5}\\\vspace{#3}\hspace{#1}{\footnotesize #4}\end{center}\end{minipage}}

\preprint{CPPC-2024-05, DESY-24-082}
\subheader{\it \today}

\title{\Large{\bf Baryon-number-violating nucleon decays in ALP effective field theories}}

\author[a]{Tong Li}
\emailAdd{litong@nankai.edu.cn}
\author[b]{Michael A. Schmidt}
\emailAdd{m.schmidt@unsw.edu.au}
\author[a,c]{Chang-Yuan Yao}
\emailAdd{yaocy@nankai.edu.cn}

\affiliation[a]{School of Physics, Nankai University, 94 Weijin Road, Tianjin 300071, China}
\affiliation[b]{
Sydney Consortium for Particle Physics and Cosmology,\\
School of Physics, The University of New South Wales, Sydney, New South Wales 2052, Australia
}
\affiliation[c]{
Deutsches Elektronen-Synchrotron DESY, Notkestr. 85, 22607 Hamburg, Germany}

\abstract{
The search for baryon-number-violating (BNV) nucleon decay is an intriguing probe of new physics beyond the SM in future neutrino experiments with enhanced sensitivity. The dark sector states such as an axion or axion-like particle (ALP) can induce nucleon decays with distinct signature and kinematics from the conventional nucleon decays.
In this work, we study the ALP effective field theories (EFTs) with baryon number violation and the impact of light ALP on BNV nucleon decays. We revisit the dimension-8 BNV operators in the extended EFTs with an ALP field $a$ respecting shift symmetry. The low-energy EFT operators with $|\Delta (B-L)|=2$ and $|\Delta (B-L)|=0$ are matched to the baryon chiral perturbation theory. We obtain the effective chiral Lagrangian and the BNV interactions between ALP and baryons/mesons. The ALP interactions lead to two-body baryon decays $B\to \ell~({\rm or}~\nu)~a$ and three-body nucleon decays $N\to M~\ell~({\rm or}~\nu)~a$. We obtain the constraints on the UV scale from the invisible $\Lambda^0$ decay search at BESIII, the invisible neutron decay search at KamLAND and proton decay search at Super-K. We also show the projections of some other baryon/nucleon decays and present the distinct distributions of kinematic observable.
}

\begin{document}

\maketitle
\setcounter{page}{2}

\newpage

\section{Introduction}
\label{sec:Intro}

The baryon number ($B$) happens to be a conserved global symmetry in the Standard Model (SM) and is only broken by non-perturbative effects. The $B$ number conservation guarantees the stability of the proton. Any observation of baryon number violation in proton decay would imply the existence of new physics (NP) beyond the SM as well as provide a necessary ingredient for the generation of the matter-antimatter asymmetry in the early Universe. The theoretical prediction of baryon-number-violating (BNV) nucleon decays has been performed in the framework of Grand Unified Theories (GUTs)~\cite{Georgi:1974sy,Fritzsch:1974nn,Langacker:1980js,Nath:2006ut}, supersymmetry~\cite{Farrar:1978xj,Dimopoulos:1981zb,Sakai:1981gr}, or the effective field theories (EFTs)~\cite{deGouvea:2014lva,Heeck:2019kgr,Girmohanta:2019xya,Antusch:2020ztu,He:2021sbl,Fajfer:2023gfi,Beneito:2023xbk,Gargalionis:2024nij,Beneke:2024hox} (see Refs.~\cite{FileviezPerez:2022ypk,Dev:2022jbf,Ohlsson:2023ddi} for recent reviews). The ongoing and future neutrino experiments, including Super-K~\cite{Takhistov:2016eqm}, Hyper-K~\cite{Hyper-Kamiokande:2018ofw}, DUNE~\cite{DUNE:2016evb,DUNE:2020ypp} and JUNO~\cite{JUNO:2015zny}, provide ideal opportunities to probe baryon number violation in nucleon decays. Super-K has searched for many nucleon decays into lepton plus meson, e.g. proton decay $p\to e^+ (\mu^+) \pi^0$, and no evidence has been found~\cite{Super-Kamiokande:2020wjk,Super-Kamiokande:2014otb,Super-Kamiokande:2017gev,Super-Kamiokande:2013rwg,Super-Kamiokande:2005lev,Super-Kamiokande:2012zik}. The lower limits on the NP energy scale can be obtained for dimension-6 or dimension-7 BNV operators in the SM effective field theory (SMEFT) and low-energy effective field theory (LEFT)~\cite{Beneito:2023xbk}.

Besides the conventional BNV search modes, nucleon decay may offer a novel
probe of dark sector particles~\cite{Davoudiasl:2014gfa,Helo:2018bgb,Barducci:2018rlx,Heeck:2020nbq,Davoudiasl:2023peu,Liang:2023yta,Fridell:2023tpb}. The new dark states with feeble interaction and negligible mass may escape the detector and mimic the standard BNV decay modes with neutrinos. The different kinematics however provides a great potential for distinction. Among the dark sector models, the axion and axion-like particle (ALP) are predicted within a wide class of NP theories~\cite{Kim:1979if,Shifman:1979if,Dine:1981rt,Zhitnitsky:1980tq,Kim:1986ax,Georgi:1986df,Kuster:2008zz} and play as good candidate of cold dark matter~\cite{Preskill:1982cy,Dine:1982ah,Abbott:1982af}. They are CP-odd pseudo-Nambu-Goldstone bosons and arise as a result of the spontaneous breaking of a global U(1) symmetry~\cite{Baluni:1978rf,Crewther:1979pi,Kim:1979if,Shifman:1979if,Dine:1981rt,Zhitnitsky:1980tq,Baker:2006ts,Pendlebury:2015lrz}. The well-studied example such as the QCD axion after breaking the Peccei-Quinn (PQ)
symmetry~\cite{Peccei:1977hh,Peccei:1977ur,Weinberg:1977ma,Wilczek:1977pj} (see Ref.~\cite{DiLuzio:2020wdo} for a recent review) would solve the strong CP problem in the SM. The ``invisible'' ALPs can be quite light with mass below the
scale of chiral symmetry breaking and a lifetime larger than the age of the universe~\cite{Georgi:1986df,Gorghetto:2018ocs}.

In this work, we explore the impact of light ALPs on BNV nucleon decays in ALP EFTs.
The operator basis of ALP EFTs with baryon/lepton number violation has recently been constructed by two groups~\cite{Song:2023lxf,Song:2023jqm,Grojean:2023tsd}.
We revisit the BNV operator basis of extended SMEFT and LEFT with an axion field $a$ respecting shift symmetry (denoted by aSMEFT and aLEFT). At mass dimension~8,
there are two BNV aSMEFT operators with $\Delta B=-1$ and $\Delta L=+1$~\cite{Song:2023lxf,Song:2023jqm,Grojean:2023tsd}
\begin{eqnarray}
\mathcal{O}_{\partial a LQd}=\epsilon^{\alpha\beta\gamma}\partial_\mu a (\bar{L}d_\alpha)(\bar{Q}^c_\beta \gamma^\mu d_\gamma)\;,~~~\mathcal{O}_{\partial a ed}=\epsilon^{\alpha\beta\gamma}\partial_\mu a (\bar{d}^c_\alpha d_\beta)(\bar{e}\gamma^\mu d_\gamma)\;,
\label{eq:aSMEFTBNV}
\end{eqnarray}
where the Greek letters denote the color indices, and the operator
$\mathcal{O}_{\partial a ed}$ only exists for $N_f \neq 1$.
After the electroweak (EW) symmetry breaking, there are eight dimension-8 aLEFT operators with $|\Delta (B-L)|=2$ and twelve with $|\Delta (B-L)|=0$. They violate both baryon and lepton numbers, and can be generated by the above two aSMEFT operators or higher-dimensional operators.
We focus on the dimension-8 aLEFT operators with $|\Delta (B-L)|=2$ or $|\Delta (B-L)|=0$ which contribute to the nucleon decay at leading order of chiral power counting.
They are matched to the baryon chiral perturbation theory (BChPT)~\cite{Claudson:1981gh,Jenkins:1990jv,Jenkins:1991ne}
~\footnote{See also Ref.~\cite{DiLuzio:2023cuk} for a recent discussion of CP-violating axion-like particles in baryon chiral perturbation theory.} based on their representations under the flavor group $SU(3)_L\times SU(3)_R$. One can then obtain the effective chiral Lagrangian at low-energies and the BNV interactions of the ALP with baryons and mesons.

These BChPT interactions induce either two-body baryon decays or three-body nucleon decays with the ALP in final states
\begin{eqnarray}
B\to \ell~({\rm or}~\nu)~a\;,~~~N\to M~\ell~({\rm or}~\nu)~a\;,~~~(\ell=e, \mu)\;,
\end{eqnarray}
where $B$, $N$ and $M$ denotes baryon, nucleon and meson, respectively. We restrict our analysis to the case of
the nearly massless and sufficiently long-lived ALP (such as the QCD axion). Their signal and kinematics are distinct from the conventional BNV nucleon decay modes $N\to M \ell~({\rm or}~\nu)$ because the invisible ALP carries away momentum.
Super-K performed the search for proton decays $p\to e^+(\mu^+)X$ with $X$ being an invisible and massless particle~\cite{Super-Kamiokande:2015pys}. KamLAND and BESIII searched for invisible decays of neutron~\cite{KamLAND:2005pen}~\footnote{There exist other search limits for the invisible decay of neutron from SNO~\cite{SNO:2003lol} and SNO+~\cite{SNO:2018ydj}. Here we take the strongest limit from KamLAND~\cite{KamLAND:2005pen}. JUNO is expected to improve the limit by two orders of magnitude~\cite{JUNO:2024pur}.} and the $\Lambda^0$ baryon~\cite{BESIII:2021slv}, respectively.
None of them found a signal and thus the limits on the partial lifetimes of proton and neutron or the invisible decay branching fraction of $\Lambda^0$ can be obtained as
\begin{eqnarray}
&&\tau(p\to e^+ (\mu^+) X)>0.79~(0.41)\times 10^{33}~{\rm years}~\text{\cite{Super-Kamiokande:2015pys}}\;,\\
&&\tau(n\to {\rm invisible})> 5.8 \times 10^{29}~{\rm years}~\text{\cite{KamLAND:2005pen}}\;,\\
&&{\rm BR}(\Lambda^0\to {\rm invisible})<7.4\times 10^{-5}~\text{\cite{BESIII:2021slv}}\;.
\end{eqnarray}
They can place upper limits on the corresponding Wilson coefficients (WCs) of ALP EFTs for certain two-body decays. We also predict the lifetimes of other nucleon decays or the decay branching fractions of other baryons.
The distributions of kinematic variable in the three-body decay modes are also shown in order to distinguish from the conventional BNV nucleon decays.
They offer promising detection of BNV nucleon/baryon decay and characteristic ALP search strategy in future neutrino experiments or BESIII upgrades with significant sensitivity.

This paper is organized as follows. In Sec.~\ref{sec:ALPEFT}, we describe the BNV operators in ALP EFTs and the BChPT. The matching of dimension-8 aLEFT operators to BChPT is discussed in details. In Sec.~\ref{sec:BNV}, we show the effective chiral Lagrangian and the baryon-number-violating interactions between ALP and baryons/mesons. The nucleon decay rates are then calculated in Sec.~\ref{sec:Results}. We present the constraints on relevant aLEFT operators and future projections. Our conclusions are drawn in Sec.~\ref{sec:Con}. The details of three-body decay matrix elements are collected in Appendix~\ref{app:3body}.

\section{ALP effective field theories}
\label{sec:ALPEFT}

In this section, we first describe the BNV EFTs for ALP~\cite{Song:2023lxf,Song:2023jqm,Grojean:2023tsd} followed by the BChPT for the calculation of nucleon decay rate.

\subsection{aSMEFT and aLEFT with baryon number violation}

Between the EW scale and some NP scale $\Lambda$, the aSMEFT describes the Hilbert series and all operators of the SMEFT extended with a light ALP field respecting shift symmetry.
The operators should be constructed in the form of a derivatively coupled ALP field $\partial_\mu a$. The baryon-number-violating operators at mass dimension 8 are $\mathcal{O}_{\partial a LQd}$ and $\mathcal{O}_{\partial a ed}$ with $|\Delta (B-L)|=2$ as shown in Eq.~(\ref{eq:aSMEFTBNV}). After the EW symmetry breaking, by specifying the flavor indices ($p,r,s,t$), we can expand these two operators as follows
\begin{eqnarray}
\mathcal{O}_{\partial a LQd,~prst}&\sim&
\epsilon^{\alpha\beta\gamma}\partial_\mu a (\bar{e}_{L,p}d_{R,\alpha r})(\bar{d}^c_{L,\beta s} \gamma^\mu d_{R,\gamma t})\label{eq:LQd1}\\
&+& \epsilon^{\alpha\beta\gamma}\partial_\mu a (\bar{\nu}_{L,p}d_{R,\alpha r})(\bar{u}^c_{L,\beta s} \gamma^\mu d_{R,\gamma t})\label{eq:LQd2}\;,\\
\mathcal{O}_{\partial a ed,~prst}&\sim& \epsilon^{\alpha\beta\gamma}\partial_\mu a (\bar{e}_{R,p}\gamma^\mu d_{R,\alpha r})(\bar{d}^c_{R,\beta s} d_{R,\gamma t})\label{eq:ed}\;.
\end{eqnarray}
Note that there is always a vector/axial-vector quark current from $\mathcal{O}_{\partial a LQd}$. The $\mathcal{O}_{\partial a ed}$ operator only involves down-type quarks and vanishes when $s=t$.

\begin{table}[htb!]
\centering
\renewcommand{\arraystretch}{1.2}
\begin{tabular}{c|c|c|c}
\hline
Name & Operator with $|\Delta (B-L)|=2$ & & $SU(3)_L\times SU(3)_R$\\
\hline
$\tilde{\mathcal{O}}^{VR,SR}_{\partial a eddd}$ & $\partial_\mu a \epsilon^{\alpha\beta\gamma}(\bar{e}_L d_{R,\alpha})(\bar{d}^c_{L,\beta}\gamma^\mu d_{R,\gamma})$~[Eq.\eqref{eq:LQd1}] &  & \\
$\mathcal{O}^{VL,SR}_{\partial a ddde}$ & $\partial_\mu a \epsilon^{\alpha\beta\gamma}(\bar{d}_{L,\alpha} \gamma^\mu d^c_{R,\beta})(\bar{d}_{L,\gamma}e_{R})$ &  \\
$\mathcal{O}^{VL,SL}_{\partial a \nu ddu}$ & $\partial_\mu a \epsilon^{\alpha\beta\gamma}(\bar{\nu}_L \gamma^\mu d_{L,\alpha})(\bar{d}^c_{L,\beta}u_{L,\gamma})$ & $\checkmark$  & $(1+8,1)$\\
$\mathcal{O}^{VR,SR}_{\partial a ud\nu d}$ & $\partial_\mu a \epsilon^{\alpha\beta\gamma}(\bar{u}^c_{L,\alpha} \gamma^\mu d_{R,\beta})(\bar{\nu}_{L}d_{R,\gamma})$~[Eq.\eqref{eq:LQd2}] & \\
$\mathcal{O}^{VR,SR}_{\partial a dd\nu u}$ & $\partial_\mu a \epsilon^{\alpha\beta\gamma}(\bar{d}^c_{L,\alpha} \gamma^\mu d_{R,\beta})(\bar{\nu}_{L}u_{R,\gamma})$ &  \\
$\mathcal{O}^{VR,SR}_{\partial a du\nu d}$ & $\partial_\mu a \epsilon^{\alpha\beta\gamma}(\bar{d}^c_{L,\alpha} \gamma^\mu u_{R,\beta})(\bar{\nu}_{L}d_{R,\gamma})$ & \\
$\mathcal{O}^{VL,SL}_{\partial a ed dd}$ & $\partial_\mu a \epsilon^{\alpha\beta\gamma}(\bar{e}_L \gamma^\mu d_{L,\alpha} )(\bar{d}^c_{L,\beta}d_{L,\gamma})$ & $\checkmark\star$ & $(8,1)$\\
$\mathcal{O}^{VR,SR}_{\partial a edd d}$ & $\partial_\mu a \epsilon^{\alpha\beta\gamma}(\bar{e}_R \gamma^\mu d_{R,\alpha})(\bar{d}^c_{R,\beta}d_{R,\gamma})$~[Eq.\eqref{eq:ed}] & $\checkmark\star$ & $(1,8)$\\
\hline
\end{tabular}
\caption{Dimension-8
aLEFT operators with $|\Delta (B-L)|=2$. The last two operators with ``$\star$'' vanish when $N_f=1$.
}
\label{tab:aLEFTBL2}
\end{table}

\begin{table}[htb!]
\centering
\renewcommand{\arraystretch}{1.2}
\begin{tabular}{c|c|c|c}
\hline
Name & Operator with $|\Delta (B-L)|=0$ & & $SU(3)_L\times SU(3)_R$\\
\hline
$\mathcal{O}^{VR,SR}_{\partial a euud}$ & $\partial_\mu a \epsilon^{\alpha\beta\gamma}(\bar{e}_L^c \gamma^\mu u_{R,\alpha})(\bar{u}_{R,\beta}^c d_{R,\gamma})$ & $\checkmark$  & $(1,8)$\\
$\mathcal{O}^{VR,SR}_{\partial a duue}$ & $\partial_\mu a \epsilon^{\alpha\beta\gamma}(\bar{d}_{L,\alpha}^c \gamma^\mu u_{R,\beta})(\bar{u}_{R,\gamma}^c e_{R})$ &  \\
$\mathcal{O}^{VL,SR}_{\partial a eudu}$ & $\partial_\mu a \epsilon^{\alpha\beta\gamma}(\bar{e}^c_R \gamma^\mu u_{L,\alpha})(\bar{d}^c_{R,\beta}u_{R,\gamma})$ &  $\checkmark$ & $(3,\bar 3)$\\
$\mathcal{O}^{VL,SR}_{\partial a dueu}$ & $\partial_\mu a \epsilon^{\alpha\beta\gamma}(\bar{d}^c_{R,\alpha} \gamma^\mu u_{L,\beta})(\bar{e}^c_{R}u_{R,\gamma})$ &   \\
$\mathcal{O}^{VL,SR}_{\partial a uude}$ & $\partial_\mu a \epsilon^{\alpha\beta\gamma}(\bar{u}_{L,\alpha} \gamma^\mu u^c_{R,\beta})(\bar{d}_{L,\gamma}e^c_{L})$ &   \\
$\mathcal{O}^{VL,SR}_{\partial a duue}$ & $\partial_\mu a \epsilon^{\alpha\beta\gamma}(\bar{d}_{L,\alpha} \gamma^\mu u^c_{R,\beta})(\bar{u}_{L,\gamma}e^c_{L})$ &  \\
$\mathcal{O}^{VL,SL}_{\partial a eudu}$ & $\partial_\mu a \epsilon^{\alpha\beta\gamma}(\bar{e}^c_{R}\gamma^\mu u_{L,\alpha})(\bar{d}^c_{L,\beta}u_{L,\gamma})$ & $\checkmark$  & $(8,1)$\\
$\mathcal{O}^{VL,SR}_{\partial a udue}$ & $\partial_\mu a \epsilon^{\alpha\beta\gamma}(\bar{u}_{L,\alpha} \gamma^\mu d^c_{R,\beta})(\bar{u}_{L,\gamma}e^c_{L})$ &   \\
$\mathcal{O}^{VL,SL}_{\partial a d\nu ud}$ & $\partial_\mu a \epsilon^{\alpha\beta\gamma}(\bar{d}^c_{R,\alpha} \gamma^\mu \nu_L)(\bar{u}^c_{L,\beta}d_{L,\gamma})$ & $\checkmark$ &$(\bar 3,3)$\\
$\mathcal{O}^{VL,SR}_{\partial a dd\nu u}$ & $\partial_\mu a \epsilon^{\alpha\beta\gamma}(\bar{d}_{L,\alpha} \gamma^\mu d^c_{R,\beta})(\bar{\nu}_{L}u^c_{L,\gamma})$ &   \\
$\mathcal{O}^{VL,SR}_{\partial a d\nu du}$ & $\partial_\mu a \epsilon^{\alpha\beta\gamma}(\bar{d}^c_{R,\alpha} \gamma^\mu \nu_{L})(\bar{d}^c_{R,\beta}u_{R,\gamma})$ & $\checkmark$ & $(1,1+8)$ \\
$\mathcal{O}^{VL,SR}_{\partial a dud\nu}$ & $\partial_\mu a \epsilon^{\alpha\beta\gamma}(\bar{d}_{L,\alpha} \gamma^\mu u^c_{R,\beta})(\bar{d}_{L,\gamma}\nu^c_{L})$ &   \\
\hline
\end{tabular}
\caption{Dimension-8
aLEFT operators with $|\Delta (B-L)|=0$.
}
\label{tab:aLEFTBL0}
\end{table}

Below the EW scale, one can also derive an operator basis for the LEFT extended with an ALP, i.e., aLEFT. The leading-order BNV operators with an ALP occur at mass dimension 8. We show the dimension-8 aLEFT operators with $|\Delta(B-L)|=2$ in Tab.~\ref{tab:aLEFTBL2} and those with $|\Delta(B-L)|=0$ in Tab.~\ref{tab:aLEFTBL0}, respectively~\footnote{We correct some errors in Table~12 of Ref.~\cite{Grojean:2023tsd} by using Fierz identity to ensure some non-vanishing operators for $N_f=1$.}.
The two operators in Tab.~\ref{tab:aLEFTBL2} with ``$\star$'' vanish when $N_f=1$.
In the following we only focus on the aLEFT operators with ``$\checkmark$'' which contribute to nucleon decay at leading order. The operators without ``$\checkmark$'' cannot be matched to BChPT at leading order of chiral power counting~\cite{Liang:2023yta}. They belong to some specific chiral irreducible representations $(6,3)$ or $(3,6)$. The matching of these subleading operators remains to be studied in details in future.
It turns out that the aSMEFT operator $\mathcal{O}_{\partial a LQd}$ can be matched to aLEFT operators $\tilde{\mathcal{O}}^{VR,SR}_{\partial a eddd}$ and $\mathcal{O}^{VR,SR}_{\partial a ud\nu d}$. However, they do not contribute to the nucleon decay at leading order.
The $\mathcal{O}_{\partial a ed}$ operator can only induce $\mathcal{O}^{VR,SR}_{\partial a eddd}$.

In Tabs.~\ref{tab:aLEFTBL2} and \ref{tab:aLEFTBL0}, we also show the transformation properties of the aLEFT operators under the quark flavor group $SU(3)_L\times SU(3)_R$. They are relevant for the matching to BChPT. The details of decomposing the tensor products in the aLEFT operators of interest are collected in App.~\ref{app:trans}.
Note that $\mathcal{O}^{VL,SL}_{\partial a \nu ddu}$ and $\mathcal{O}^{VL,SR}_{\partial a d\nu du}$ operators also induce the trivial representation $(1,1)$. The singlet-flavor state identically vanishes for the ground-state baryon due to Fermi statistics and thus we do not take it into account in the following. The singlet state in flavor space does not exist for the ground state with zero angular momentum due to Fermi statistics. The reason is as follows. The wave function consists of the spatial, spin, flavor, and color part. The spatial part of the ground state is symmetric because of having zero angular momentum. The flavor and color parts are both fully anti-symmetric, but the spin wave function is not fully anti-symmetric, which is required to obtain a fully anti-symmetric wave function.

\subsection{Matching to the BChPT}

In this section, we match the relevant aLEFT operators to BChPT and derive the ALP interactions with baryon and meson. The nuclear matrix elements are also given for the calculation of nucleon decay rates. The chiral Lagrangian based on chiral $SU(3)_L \times SU(3)_R$ symmetry has been proved useful in describing low-energy hadronic physics. The power counting expansion in the interactions of the chiral Lagrangian is valid for processes at low energy. Here we consider the strong interactions of the pseudo-Goldstone mesons with the baryon fields in $B$ octet, which was first constructed by Claudson, Wise and Hall in Ref.~\cite{Claudson:1981gh}. It is important to note that the decaying baryons are non-relativistic and can be treated as heavy fermions. Thus, it is possible to implement a derivative expansion in chiral perturbation theory. This was discussed in Refs.~\cite{Jenkins:1990jv,Jenkins:1991ne}. Although the BChPT is currently the best available technique for
the calculation of nucleon decay, this approach has limitations as it works best for small meson momenta~\cite{Nath:2006ut}. We justify the use of BChPT for the considered processes in Sec.~\ref{sec:BNV}.

The baryon $B$, $\overline{B}$ and meson $M$ fields in BChPT are given by
\begin{eqnarray}
&&B=\left(
  \begin{array}{ccc}
    {\Sigma^0\over \sqrt{2}}+{\Lambda^0\over \sqrt{6}} & \Sigma^+ & p \\
    \Sigma^- & -{\Sigma^0\over \sqrt{2}}+{\Lambda^0\over \sqrt{6}} & n \\
    \Xi^- & \Xi^0 & -\sqrt{2\over 3}\Lambda^0 \\
  \end{array}
\right)\;,~~\overline{B}=\left(
  \begin{array}{ccc}
    {\overline{\Sigma^0}\over \sqrt{2}}+{\overline{\Lambda^0}\over \sqrt{6}} & \overline{\Sigma^-} & \overline{\Xi^-} \\
    \overline{\Sigma^+} & -{\overline{\Sigma^0}\over \sqrt{2}}+{\overline{\Lambda^0}\over \sqrt{6}} & \overline{\Xi^0} \\
    \overline{p} & \overline{n} & -\sqrt{2\over 3}\overline{\Lambda^0} \\
  \end{array}
\right)\;,\\
&&M=\left(
  \begin{array}{ccc}
    {\pi^0\over \sqrt{2}}+{\eta^0\over \sqrt{6}} & \pi^+ & K^+ \\
    \pi^- & -{\pi^0\over \sqrt{2}}+{\eta^0\over \sqrt{6}} & K^0 \\
    K^- & \bar{K}^0 & -\sqrt{2\over 3}\eta^0 \\
  \end{array}
\right)\;,~~M^\dagger=M\;,~~\Sigma\equiv e^{2iM/f_\pi}\;,~~\xi\equiv e^{iM/f_\pi}\;.
\end{eqnarray}
In principle, the neutral components of the baryon or meson octet should mix to form the physical states after isospin breaking. Here we ignore the quark mass splitting effects both for baryons and mesons because they are generally small. The leading-order chiral Lagrangian for baryons and mesons is then given by~\cite{Claudson:1981gh}
\begin{eqnarray}
\mathcal{L}={1\over 8}f_\pi^2 {\rm tr}(\partial_\mu \Sigma \partial^\mu \Sigma^\dagger) + {\rm tr}(\overline{B}(i\cancel{D}-M_B)B)-{D\over 2}{\rm tr}(\overline{B}\gamma^\mu \gamma_5 \{\xi_\mu,B\})-{F\over 2}{\rm tr}(\overline{B}\gamma^\mu \gamma_5 [\xi_\mu,B])\;,
\label{eq:ChPT}
\end{eqnarray}
where $D_\mu B=\partial_\mu B + [\Gamma_\mu,B]$ with
\begin{eqnarray}
\Gamma_\mu={1\over 2}(\xi^\dagger \partial_\mu \xi+\xi \partial_\mu \xi^\dagger)\;,~~~\xi_\mu=i(\xi^\dagger \partial_\mu \xi-\xi \partial_\mu \xi^\dagger)\;.
\end{eqnarray}
It gives the couplings of two baryons to a meson as shown in Refs.~\cite{Claudson:1981gh,Nath:2006ut,Beneito:2023xbk}.

To match the aLEFT operators to BChPT, we use the flavor symmetry and match the operators which have the same flavor symmetry properties~\cite{Claudson:1981gh}. The flavor symmetry transformations of the relevant operators in BChPT are
\begin{align}
    \xi B \xi \sim (3,\bar 3)& \to L \xi B \xi R^\dagger \;, &
    \xi^\dagger B \xi^\dagger  \sim (\bar 3,3) & \to R \xi^\dagger B \xi^\dagger L^\dagger \;,
    \\\nonumber
    \xi B \xi^\dagger \sim(8,1) & \to L \xi B \xi^\dagger L^\dagger \;, &
    \xi^\dagger B \xi \sim (1,8)& \to R \xi^\dagger B \xi R^\dagger \;,
\end{align}
where $L$ ($R$) is an element of $SU(3)_L$ ($SU(3)_R$).
In addition, we utilize that the leading-order nuclear matrix elements can be related to the two independent low-energy constants $\alpha$ and $\beta$ which are defined by
\begin{eqnarray}
&&\langle 0|\epsilon^{\alpha\beta\gamma} (\bar{u}^c_{R,\alpha} d_{R,\beta}) u_{L,\gamma}|p^{(s)}\rangle = \alpha u^{(s)}_{pL}\;,~~\langle 0|\epsilon^{\alpha\beta\gamma} (\bar{u}^c_{L,\alpha} d_{L,\beta}) u_{L,\gamma}|p^{(s)}\rangle = \beta u^{(s)}_{pL}\;.
\end{eqnarray}
The matrix elements of other baryons in the baryon octet can be related to those two by imposing SU(3) flavour symmetry and requiring parity conservation
\begin{eqnarray}
&&\langle 0|\epsilon^{\alpha\beta\gamma} (\bar{q}^c_{R,i\alpha} q_{R,j\beta}) q_{L,k\gamma}|B^{(s)}\rangle = \alpha u^{(s)}_{BL}\;,
~~\langle 0|\epsilon^{\alpha\beta\gamma} (\bar{q}^c_{L,i\alpha} q_{L,j\beta}) q_{L,k\gamma}|B^{(s)}\rangle = \beta u^{(s)}_{BL}\;,
\\
&&\langle 0|\epsilon^{\alpha\beta\gamma} (\bar{q}^c_{L,i\alpha} q_{L,j\beta}) q_{R,k\gamma}|B^{(s)}\rangle = -\alpha u^{(s)}_{BR}\;,
~~\langle 0|\epsilon^{\alpha\beta\gamma} (\bar{q}^c_{R,i\alpha} q_{R,j\beta}) q_{R,k\gamma}|B^{(s)}\rangle = -\beta u^{(s)}_{BR}\;.
\end{eqnarray}

In the following expansion, we only keep relevant terms for the calculation of BNV nucleon decay at leading order.
The terms at order $f_\pi^0$ are complete and they induce all possible baryons/nucleons' two-body decays. At order $f_\pi^{-1}$, we only keep the terms with protons and neutrons which give nucleons' three-body decays.
There are indeed other baryons' couplings at order $f_\pi^{-1}$ which induce their three-body decays. However, as we will explain in next section, other baryons' three-body decays would not be as competitive as either their two-body decays or the three-body decays of nucleons that we consider in this work. Thus, we drop terms with other baryons at order $f_\pi^{-1}$.
The projection matrices for each operator can be read off from the decompositions in App.~\ref{app:trans}.
The projection operator $P_{ij}$ ($\tilde{P}_{ij}$) is given by setting the $(i,j)$ entry to $1 \; (-1)$ and all other entries to zero.
The matching of $|\Delta(B-L)|=2$ aLEFT operators to BChPT becomes
\begin{align}
[\mathcal{O}^{VL,SL}_{\partial a \nu ddu}]_{r221} &\to \beta \partial_\mu a \bar{\nu}_{L} \gamma^\mu {\rm tr}(\xi B \xi^\dagger \tilde{P}_{32}) \nonumber \\
&\supset \partial_\mu a \bar \nu_L \gamma^\mu [-\beta n + \frac{i\beta}{f_\pi}(-\sqrt{\frac32}n\eta +\frac{1}{\sqrt{2}}n\pi^0-p \pi^-)] \;,\\
[\mathcal{O}^{VL,SL}_{\partial a \nu ddu}]_{r231} &\to \beta \partial_\mu a \bar{\nu}_{L} \gamma^\mu {\rm tr}(\xi B \xi^\dagger ({2\over 3}P_{22}+{1\over 3}\tilde{P}_{11}+{1\over 3}\tilde{P}_{33}))\nonumber \\
&\supset \partial_\mu a \bar{\nu}_{L} \gamma^\mu [\beta {1\over \sqrt{6}}\Lambda^0-\beta {1\over \sqrt{2}}\Sigma^0 -{i\beta\over f_\pi} \bar{K}^0n]\;,\\
[\mathcal{O}^{VL,SL}_{\partial a \nu ddu}]_{r321} &\to \beta \partial_\mu a \bar{\nu}_{L} \gamma^\mu {\rm tr}(\xi B \xi^\dagger ({1\over 3}P_{22}+{1\over 3}P_{11}+{2\over 3}\tilde{P}_{33}))\nonumber \\
&\supset \partial_\mu a \bar{\nu}_{L} \gamma^\mu [\beta \sqrt{2\over 3}\Lambda^0 -{i\beta\over f_\pi} (\bar{K}^0n+K^- p)]\;,\\
[\mathcal{O}^{VL,SL}_{\partial a \nu ddu}]_{r331} &\to \beta \partial_\mu a \bar{\nu}_{L} \gamma^\mu {\rm tr}(\xi B \xi^\dagger P_{23})\supset   \partial_\mu a \bar \nu_L \gamma^\mu [\beta \Xi^0]  \;.
\end{align}
\begin{eqnarray}
&&[\mathcal{O}^{VL,SL}_{\partial a eddd}]_{r223} \to \beta \partial_\mu a \bar{e}_{L} \gamma^\mu {\rm tr}(\xi B \xi^\dagger P_{12})\supset \partial_\mu a \bar e_L \gamma^\mu [\beta \Sigma^-  - \frac{i\beta}{f_\pi}n K^-]\;,\\
&&[\mathcal{O}^{VL,SL}_{\partial a eddd}]_{r323} \to \beta \partial_\mu a \bar{e}_{L} \gamma^\mu {\rm tr}(\xi B \xi^\dagger P_{13})\supset  \partial_\mu a \bar e_L \gamma^\mu [\beta \Xi^-] \;.
\end{eqnarray}
\begin{eqnarray}
&&[\mathcal{O}^{VR,SR}_{\partial a eddd}]_{r223} \to -\beta \partial_\mu a \bar{e}_{R} \gamma^\mu {\rm tr}(\xi^\dagger B \xi P_{12})\supset \partial_\mu a \bar e_R \gamma^\mu [-\beta \Sigma^- - \frac{i\beta}{f_\pi} n K^-]\;,\\
&&[\mathcal{O}^{VR,SR}_{\partial a eddd}]_{r323} \to -\beta \partial_\mu a \bar{e}_{R} \gamma^\mu {\rm tr}(\xi^\dagger B \xi P_{13})\supset \partial_\mu a \bar e_R \gamma^\mu [-\beta \Xi^- ] \;.
\end{eqnarray}

For the $|\Delta(B-L)|=0$ aLEFT operators,
we make use of the following identities to rewrite some operators
\begin{eqnarray}
\bar{\psi}^c_a P_{L,R} \psi_b = \bar{\psi}^c_b P_{L,R} \psi_a\;,~~\bar{\psi}^c_a \gamma^\mu P_{L,R} \psi_b = -\bar{\psi}^c_b \gamma^\mu P_{R,L} \psi_a\;,
\end{eqnarray}
where $a$ and $b$ include all indices of the fields (color, flavor, and etc.).
Then, we have
\begin{eqnarray}
&&\mathcal{O}^{VL,SR}_{\partial a eudu}=\partial_\mu a \epsilon^{\alpha\beta\gamma}(\bar{e}^c_{R} \gamma^\mu u_{L,\alpha})(\bar{d}^c_{R,\beta}u_{R,\gamma})=-\partial_\mu a \epsilon^{\alpha\beta\gamma}(\bar{e}^c_{R} \gamma^\mu u_{L,\alpha})(\bar{u}^c_{R,\beta}d_{R,\gamma})\;,\\
&&\mathcal{O}^{VL,SL}_{\partial a eudu}=\partial_\mu a \epsilon^{\alpha\beta\gamma}(\bar{e}^c_{R} \gamma^\mu u_{L,\alpha})(\bar{d}^c_{L,\beta}u_{L,\gamma})=-\partial_\mu a \epsilon^{\alpha\beta\gamma}(\bar{e}^c_{R} \gamma^\mu u_{L,\alpha})(\bar{u}^c_{L,\beta}d_{L,\gamma})\;,\\
&&\mathcal{O}^{VL,SL}_{\partial a d\nu ud}=\partial_\mu a \epsilon^{\alpha\beta\gamma}(\bar{d}^c_{R,\alpha}\gamma^\mu \nu_L)(\bar{u}^c_{L,\beta}d_{L,\gamma})=\partial_\mu a \epsilon^{\alpha\beta\gamma}(\bar{\nu}^c_L\gamma^\mu d_{R,\alpha})(\bar{d}^c_{L,\beta}u_{L,\gamma})\;,\\
&&\mathcal{O}^{VL,SR}_{\partial a d\nu du}=\partial_\mu a \epsilon^{\alpha\beta\gamma}(\bar{d}^c_{R,\alpha}\gamma^\mu \nu_L)(\bar{d}^c_{R,\beta}u_{R,\gamma})=-\partial_\mu a \epsilon^{\alpha\beta\gamma}(\bar{\nu}^c_{L}\gamma^\mu d_{R,\alpha})(\bar{d}^c_{R,\beta}u_{R,\gamma})\;.
\end{eqnarray}
The matching of $|\Delta(B-L)|=0$ aLEFT operators to BChPT becomes
\begin{align}
[\mathcal{O}^{VR,SR}_{\partial a euud}]_{r 112} &\to -\beta \partial_\mu a \bar{e}^c_{L} \gamma^\mu {\rm tr}(\xi^\dagger B \xi P_{31})\nonumber\\
&\supset \partial_\mu a \bar{e}^c_{L} \gamma^\mu [-\beta p + {i\beta\over f_\pi}(\sqrt{3\over 2}p\eta+{1\over \sqrt{2}}p\pi^0+n\pi^+)]\;,\\
[\mathcal{O}^{VR,SR}_{\partial a euud}]_{r 113} &\to -\beta \partial_\mu a \bar{e}^c_{L} \gamma^\mu {\rm tr}(\xi^\dagger B \xi \tilde{P}_{21})\supset \partial_\mu a \bar{e}^c_{L} \gamma^\mu [\beta \Sigma^+ + {i\beta\over f_\pi}p\bar{K}^0]\;.
\end{align}
\begin{align}
[\mathcal{O}^{VL,SR}_{\partial a eudu}]_{r121} &\to -\alpha \partial_\mu a \bar{e}^c_{R} \gamma^\mu {\rm tr}(\xi B \xi P_{31})\nonumber \\
&\supset \partial_\mu a \bar{e}^c_{R} \gamma^\mu [-\alpha p - {i\alpha\over f_\pi}(-{1\over\sqrt{6}}p\eta+{1\over \sqrt{2}}p\pi^0+n\pi^+)]\;,\\
[\mathcal{O}^{VL,SR}_{\partial a eudu}]_{r131} &\to -\alpha \partial_\mu a \bar{e}^c_{R} \gamma^\mu {\rm tr}(\xi B \xi \tilde{P}_{21})\supset \partial_\mu a \bar{e}^c_{R} \gamma^\mu [\alpha \Sigma^+ + {i\alpha\over f_\pi}p\bar{K}^0]\;.
\end{align}
\begin{align}
[\mathcal{O}^{VL,SL}_{\partial a eudu}]_{r121} &\to -\beta \partial_\mu a \bar{e}^c_{R} \gamma^\mu {\rm tr}(\xi B \xi^\dagger P_{31})\nonumber \\
&\supset \partial_\mu a \bar{e}^c_{R} \gamma^\mu [-\beta p - {i\beta\over f_\pi}(\sqrt{3\over 2}p\eta+{1\over \sqrt{2}}p\pi^0+n\pi^+)]\;,\\
[\mathcal{O}^{VL,SL}_{\partial a eudu}]_{r131} &\to -\beta \partial_\mu a \bar{e}^c_{R} \gamma^\mu {\rm tr}(\xi B \xi^\dagger \tilde{P}_{21})\supset \partial_\mu a \bar{e}^c_{R} \gamma^\mu [\beta \Sigma^+ - {i\beta\over f_\pi}p\bar{K}^0]\;.
\end{align}
\begin{align}
[\mathcal{O}^{VL,SL}_{\partial a d\nu ud}]_{2r12} &\to -\alpha \partial_\mu a \bar{\nu}^c_{L} \gamma^\mu {\rm tr}(\xi^\dagger B \xi^\dagger \tilde{P}_{32})\nonumber \\
&\supset \partial_\mu a \bar{\nu}^c_{L} \gamma^\mu [\alpha n + {i\alpha\over f_\pi}({1\over\sqrt{6}}n\eta+{1\over \sqrt{2}}n\pi^0-p\pi^-)]\;,\\
[\mathcal{O}^{VL,SL}_{\partial a d\nu ud}]_{2r13} &\to -\alpha \partial_\mu a \bar{\nu}^c_{L} \gamma^\mu {\rm tr}(\xi^\dagger B \xi^\dagger P_{22})\supset \partial_\mu a \bar{\nu}^c_{L} \gamma^\mu [\alpha ({1\over \sqrt{2}} \Sigma^0 - {1\over \sqrt{6}}\Lambda^0) + {i\alpha\over f_\pi} n \bar{K}^0]\;,\\
[\mathcal{O}^{VL,SL}_{\partial a d\nu ud}]_{3r12} &\to -\alpha \partial_\mu a \bar{\nu}^c_{L} \gamma^\mu {\rm tr}(\xi^\dagger B \xi^\dagger \tilde{P}_{33})\supset \partial_\mu a \bar{\nu}^c_{L} \gamma^\mu [-\alpha \sqrt{2\over 3}\Lambda^0 - {i\alpha\over f_\pi} (n\bar{K}^0+pK^-)]\;,\\
[\mathcal{O}^{VL,SL}_{\partial a d\nu ud}]_{3r13} &\to -\alpha \partial_\mu a \bar{\nu}^c_{L} \gamma^\mu {\rm tr}(\xi^\dagger B \xi^\dagger P_{23})\supset \partial_\mu a \bar{\nu}^c_{L} \gamma^\mu [-\alpha \Xi^0]\;.
\end{align}
\begin{align}
[\mathcal{O}^{VL,SR}_{\partial a d\nu du}]_{2r21} &\to \beta \partial_\mu a \bar{\nu}^c_{L} \gamma^\mu {\rm tr}(\xi^\dagger B \xi \tilde{P}_{32})\nonumber \\
&\supset  \partial_\mu a \bar \nu_L^c \gamma^\mu [  -\beta n +\frac{i\beta}{f_\pi} (\sqrt{\frac32} n\eta -\frac{1}{\sqrt{2}}n\pi^0+p \pi^-)]\;,\\
[\mathcal{O}^{VL,SR}_{\partial a d\nu du}]_{2r31} &\to \beta \partial_\mu a \bar{\nu}^c_{L} \gamma^\mu {\rm tr}(\xi^\dagger B \xi ({2\over 3}P_{22}+{1\over 3}\tilde{P}_{11}+{1\over 3}\tilde{P}_{33}))\nonumber \\
&\supset \partial_\mu a \bar{\nu}^c_{L} \gamma^\mu [\beta {1\over \sqrt{6}}\Lambda^0-\beta {1\over \sqrt{2}}\Sigma^0 +{i\beta\over f_\pi} \bar{K}^0n]\;,\\
[\mathcal{O}^{VL,SR}_{\partial a d\nu du}]_{3r21} &\to \beta \partial_\mu a \bar{\nu}^c_{L} \gamma^\mu {\rm tr}(\xi^\dagger B \xi ({1\over 3}P_{22}+{1\over 3}P_{11}+{2\over 3}\tilde{P}_{33}))\nonumber\\
&\supset \partial_\mu a \bar{\nu}^c_{L} \gamma^\mu [\beta \sqrt{2\over 3}\Lambda^0 +{i\beta\over f_\pi} (\bar{K}^0n+K^- p)]\;,\\
[\mathcal{O}^{VL,SR}_{\partial a d\nu du}]_{3r31} &\to \beta \partial_\mu a \bar{\nu}^c_{L} \gamma^\mu {\rm tr}(\xi^\dagger B \xi P_{23})\supset \partial_\mu a \bar \nu_L^c \gamma^\mu [\beta\Xi^0]\;.
\end{align}

\subsection{Discussions on UV model}

As a phenomenological study, we remain agnostic about the origin of the ALP BNV operators. The WCs of ALP BNV operators were simply taken as independent parameters in an effective framework. Nevertheless, it is realized that the SM fields and their chiralities in aSMEFT operators may be quite different from the BNV operators in SMEFT. One can make use of this property to build a BNV aSMEFT operator and meanwhile forbid BNV SMEFT operators to avoid strong constraints from conventional nucleon decays.

For illustration, we provide a UV model to induce the dimension-8 operator $\mathcal{O}_{\partial a ed}$ after integrating out the new particles.
Note that operator $\mathcal{O}_{\partial a ed}$ consists of the right-handed down-type quarks $d$ and charged leptons $e$ in addition to the derivative coupling to the axion field. It is not possible to construct any BNV SMEFT operator out of $d$ and $e$ of the same or lower dimension. This provides an opportunity to obtain $\mathcal{O}_{\partial aed}$ without simultaneously inducing a more relevant BNV SMEFT operator.

\begin{figure}[tb!]
\centering
\includegraphics[width=0.5\textwidth]{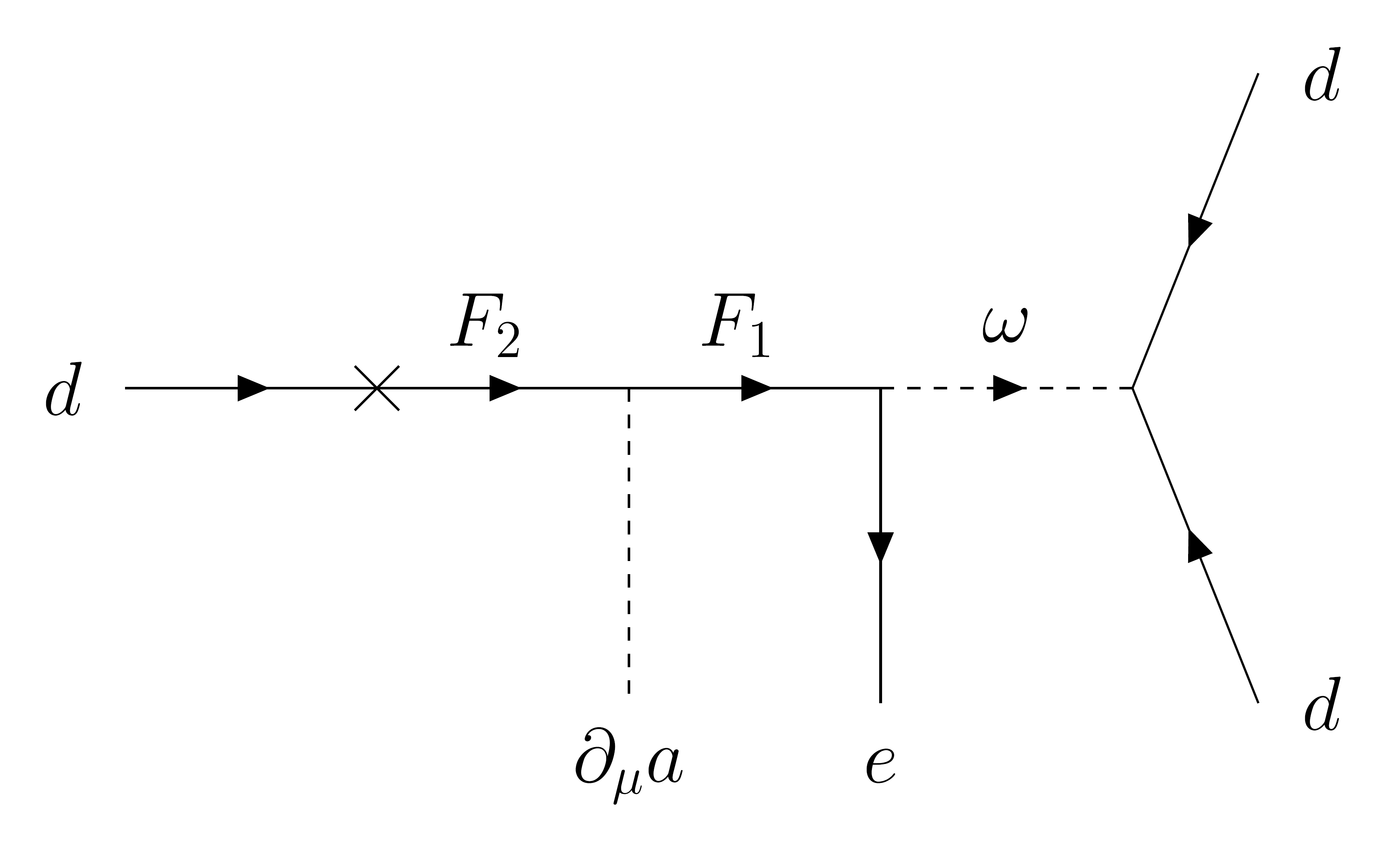}
\caption{The Feynman diagram for the UV realization of dimension-8 aSMEFT operator $\mathcal{O}_{\partial a ed}$. The diagrams here and below are generated using the TikZ-Feynman and TikZ-FeynHand packages~\cite{Ellis:2016jkw,Dohse:2018vqo}.
}
\label{fig:UV}
\end{figure}

We introduce a scalar diquark $\omega\sim (3,1,2/3)$~\footnote{The conjugate of the diquark $\bar S_1$ in the notation of Refs.~\cite{Buchmuller:1986zs,Dorsner:2016wpm}.} under SM groups $SU(3)_c\times SU(2)_L\times U(1)_Y$ and two vector-like fermions $F_i=F_{iL}+F_{iR}\sim (3,1,-1/3)$ with $i=1,2$. The diquark $\omega$ has a coupling with right-handed down-type quarks. The key ingredient however to obtain $\mathcal{O}_{\partial aed}$ without inducing a relevant BNV SMEFT operator is the flavor structure of the vector-like fermion couplings. We choose $F_2$ to mix with the SM down-type quarks, while $F_1$ features an interaction with a right-handed charged lepton and  $\omega$.
We further consider a flavor-violating ALP coupling to the vector-like fermions. The relevant part of the effective Lagrangian thus becomes
\begin{eqnarray}
\mathcal{L} =
y_{1,ij}\omega \bar{d}^c_i d_j + m_{Fd,k} \overline{F_{2L}}d_k + y_{2,l}\omega^\dagger \bar{e}_l F_{1L}
+{\partial_\mu a \over \Lambda} \overline{F_{1L}} C_{12} \gamma^\mu F_{2L}
+ {\rm h.c.}\;.
\end{eqnarray}
The Feynman diagram of the relevant process is shown in Fig.~\ref{fig:UV}. After integrating out the heavy particles $F_i$ and $\omega$, one can generate the $\mathcal{O}_{\partial a ed}$ operator and the WC is given by
\begin{eqnarray}
C_{\partial a ed,prst} = y_{1,st} y_{2,p} {C_{12}m_{Fd,r}\over m_\omega^2 m_{F_1}m_{F_2}\Lambda}\;,
\end{eqnarray}
where $m_\omega$ and $m_{F_i}$ denote the diquark and vector-like fermion masses. This operator changes $B-L$ by 2 units ($|\Delta (B-L)|=2$) and contributes to the two-body decays $\Sigma^- \to \ell^-a$, $\Xi^-\to \ell^-a$ and the three-body neutron decay $n\to K^+\ell^-a$ with $\ell=e,\mu$.

Integrating out $\omega$ and the vector-like quarks $F_i$ results in additional operators at tree level. The interaction of $\omega$ with 2 down-type quarks results in an operator with 4 down-type quarks. As the Yukawa coupling $y_1$ is antisymmetric, the relevant operator is $(\bar s\gamma^\mu P_R s)(\bar d\gamma_\mu P_R d)$ which is flavor conserving and hence it does not contribute to meson mixing. It contributes to decays of flavorless mesons with strange quarks and diquark production at the LHC, which do not yield any stringent constraints since they are dominated by strong interactions at tree level. If $\omega$ also couples to the bottom quark, there is a contribution to the $B\to K$ flavor-violating processes. As $F_2$ mixes with right-handed down-type quarks, integrating out $F_2$ results in a correction to the kinetic term of the right-handed down-type quarks at leading order which can be reabsorbed in the definition of the down-type quark Yukawa coupling. No relevant BNV SMEFT operators are induced due to the flavor structure of the vector-like fermions $F_i$.

\section{Baryon-number-violating Lagrangian and baryon decay rates}
\label{sec:BNV}

\begin{figure}[tb!]
\centering
\includegraphics[width=0.35\textwidth]{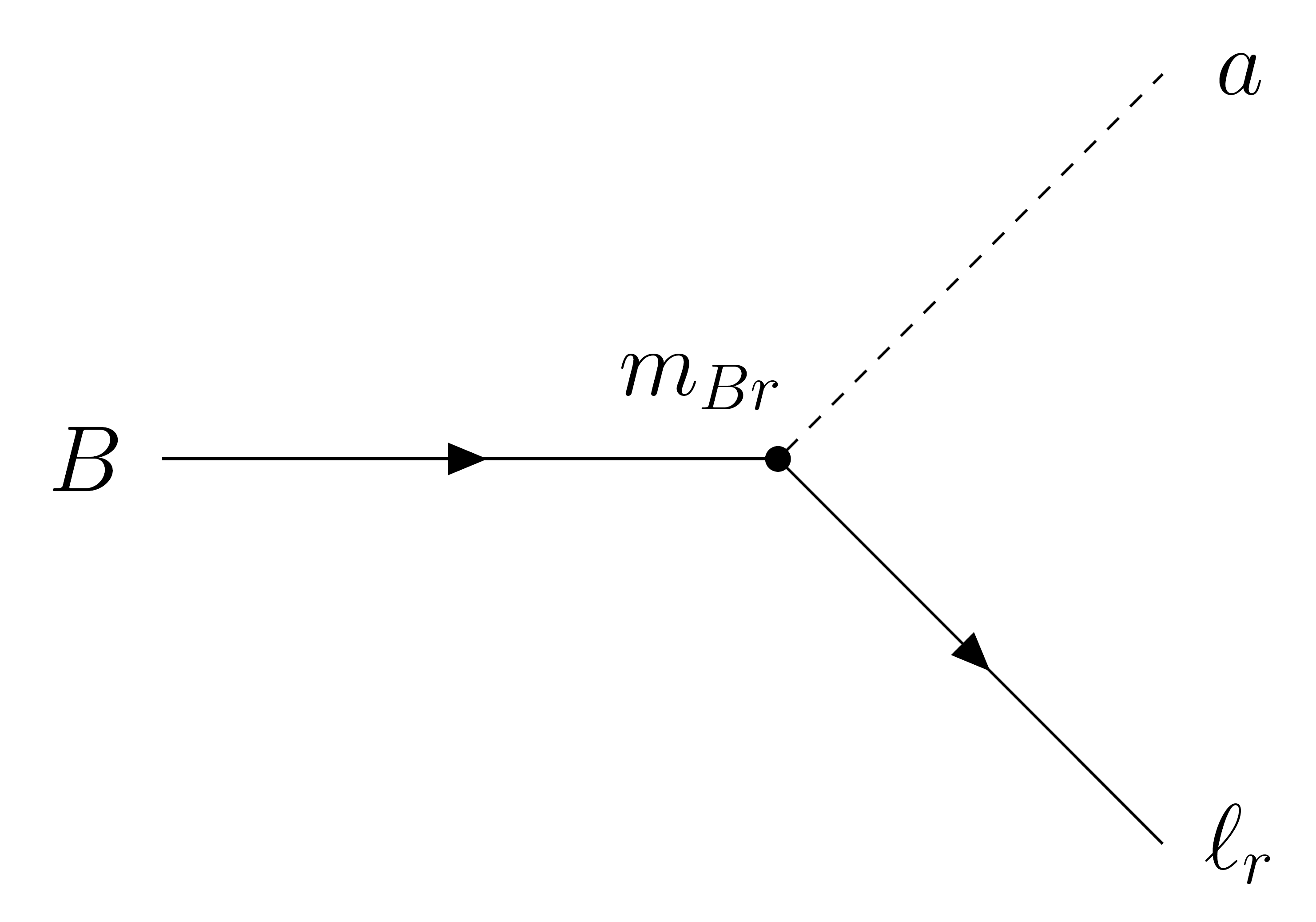}\\
\includegraphics[width=0.35\textwidth]{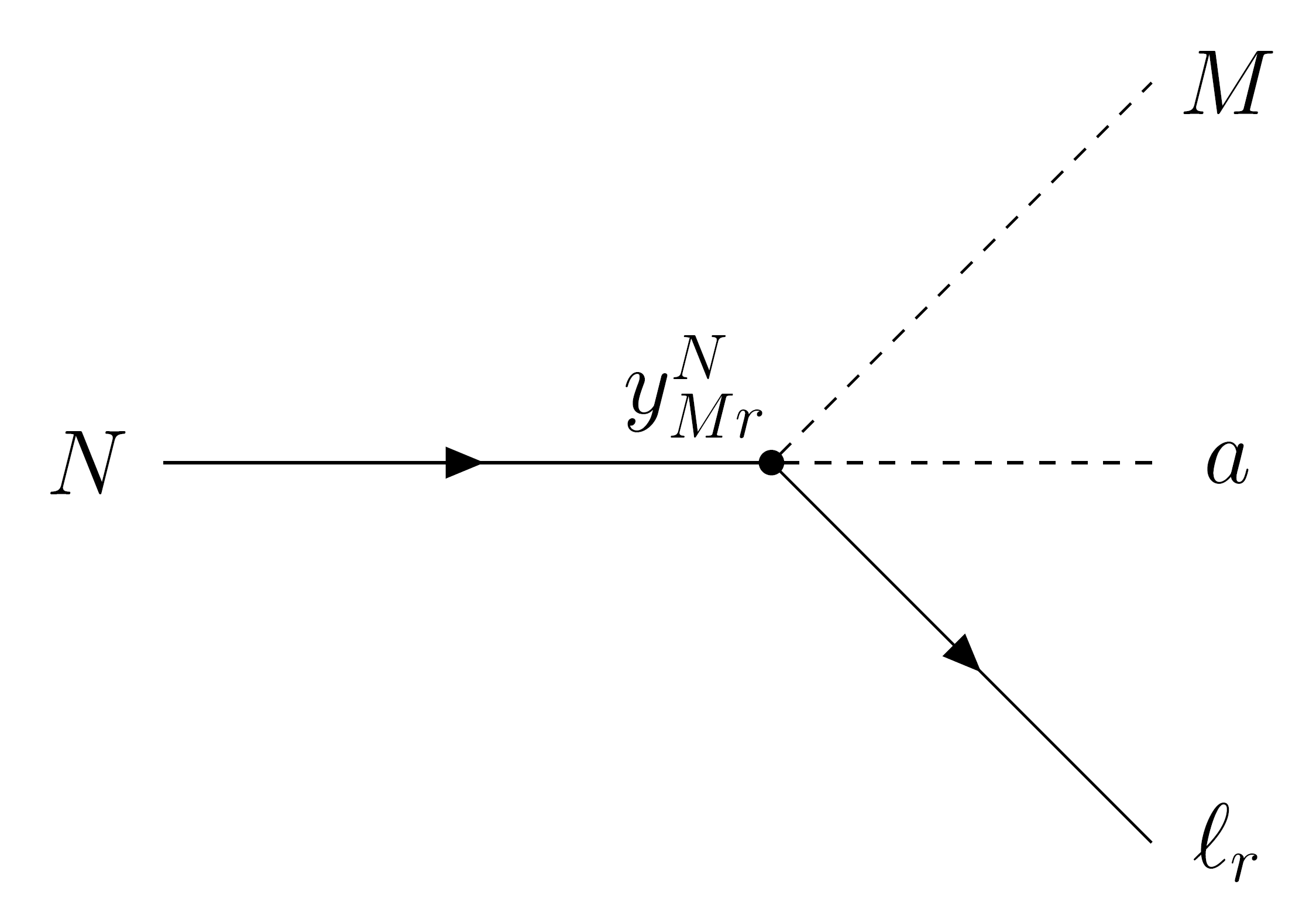}
\includegraphics[width=0.35\textwidth]{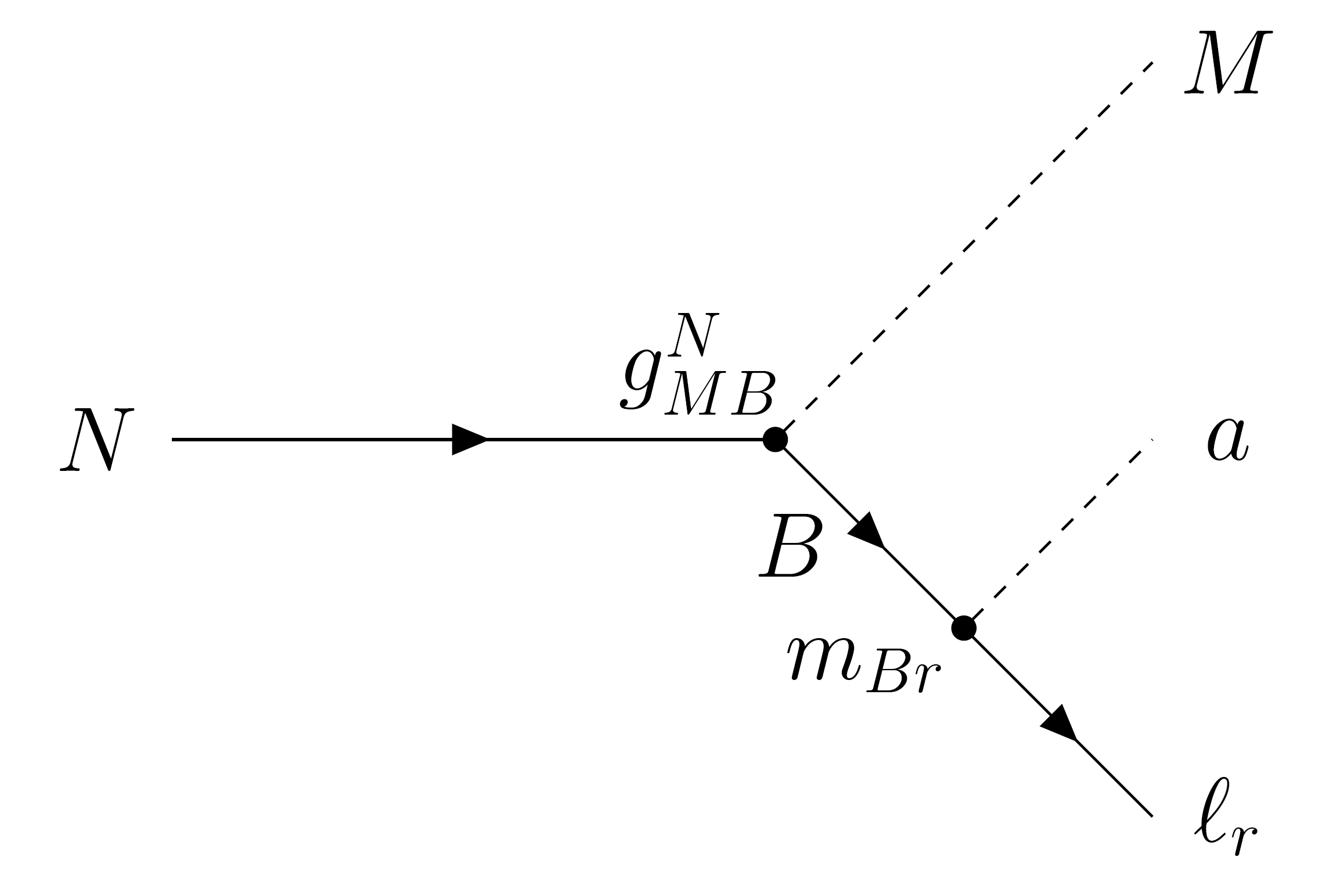}
\caption{The Feynman diagrams for the two-body baryon decay $B\to \ell a$ (top) and the three-body nucleon decay $N\to M \ell a$ (bottom).
}
\label{fig:Feynman}
\end{figure}

The $|\Delta(B-L)|=2$ effective Lagrangian of relevant interactions is
\begin{eqnarray}
\mathcal{L}=g_{MB}^N \overline{B} \gamma^\mu \gamma_5 N \partial_\mu M + m_{Br,X}\partial_\mu a \overline{\ell_r} P_{\bar{X}}\gamma^\mu B + iy_{Mr,X}^N \partial_\mu a \overline{\ell_r} P_{\bar{X}}\gamma^\mu N M \;,
\end{eqnarray}
where $X=L, R$ denotes the chirality of lepton $\ell_r$ ($\ell_1=e$, $\ell_2=\mu$), and $\bar{X}$ is the opposite chirality with $\bar{L}=R$ and $\bar{R}=L$.
The couplings $g_{MB}^N$, $m_{Br,X}$ and $y_{Mr,X}^N$ can be read from the chiral Lagrangian Eq.~(\ref{eq:ChPT}) and the above results of matching aLEFT operators to BChPT, respectively. We collect the necessary couplings in Tabs.~\ref{tab:2bodyBL2}, \ref{tab:2bodyBL0}, \ref{tab:3bodyBL2} and \ref{tab:3bodyBL0}.
In Fig.~\ref{fig:Feynman}, we show the Feynman diagrams for the two-body baryon decay $B\to \ell a$ (top) and the three-body nucleon decay $N\to M \ell a$ (bottom).
Note that in principle there exist three-body decay processes for various baryons besides nucleons. However, for other baryons, their three-body decays would not be as competitive as either their two-body decays in our Tabs.~\ref{tab:2bodyBL2} and \ref{tab:2bodyBL0} or the three-body decays of nucleons in Tabs.~\ref{tab:3bodyBL2} and \ref{tab:3bodyBL0}. Other baryons' three-body decays suffer from phase space suppression, compared to their own two-body decays. Moreover, the upcoming neutrino experiments would significantly enhance their sensitivity to nucleon decay by about one order of magnitude. The nucleon decay provides a promising probe of dark particles such as ALP. Thus, we emphasize the nucleon three-body decays to meson, lepton and ALP.

For the conventional nucleon decay into a meson and a lepton, the momentum of the final-state meson may approach as large as $m_N/2\sim 500$ MeV $\sim \Lambda_{\rm QCD}$ and thus reach the limit of chiral perturbation theory. Nevertheless, the nuclear matrix element for a two-body decay to an ALP and lepton in our case is given by the nucleon decay constants $\alpha$ and $\beta$ which do not assume a small momentum. For the three-body nucleon decays, the ALP carries away momentum and thus the meson tends to move towards the smaller momentum region ($<500$ MeV) as we show in our Figs.~\ref{fig:PM_BL2} and \ref{fig:PM_BL0}, which improves the reliability of the result.

In the SMEFT, higher-dimensional operators may induce similar signature to ours. There is a dimension-9 operator $\mathcal{O}_{LLLQuu}=\epsilon_{\alpha\beta\gamma}\epsilon_{ik}\epsilon_{jl}(\overline{L^{c,i}}L^j)(\overline{L^{c,k}}Q^{l,\alpha})(\overline{u^{c,\beta}}u^{\gamma})$ with $(\Delta B,\Delta L)=(1,3)$~\cite{Liao:2020jmn,Li:2020xlh} that may induce the four-body nucleon decay to a meson, a lepton and two neutrinos. The four-body decays suffers from a phase-space suppression but may mimic our case of three-body nucleon decay to meson, lepton and a nearly massless ALP. However, because of the bilinear $\epsilon_{\alpha\beta\gamma}\overline{u^{c,\beta}}u^{\gamma}$, this operator is present when $N_f\neq 1$ and is composed of at least two up-type quark flavors. Thus, this operator doesn't induce nucleon decay and is irrelevant to our study. In the LEFT, there exist additional relevant operators~\cite{Li:2020tsi}. Nevertheless, they are from higher-dimensional SMEFT operators which have different Lorentz structures and are not produced by our ad-hoc model.

The matrix element of two-body decay $B\to \ell a$ then becomes
\begin{eqnarray}
i\mathcal{M}=\bar{u}_{\ell_r} \sum_{X=L,R} P_{\bar{X}}(-m_{Br,X}\cancel{p}_a) u_B\;,
\end{eqnarray}
where $p_a$ denotes the outgoing ALP momentum in the final state.
The matrix element of $N\to M \ell a$ is
\begin{eqnarray}
i\mathcal{M}=\bar{u}_{\ell_r} \sum_{X=L,R} P_{\bar{X}}(-iy_{Mr,X}^N \cancel{p}_a + \sum_B m_{Br,X}\cancel{p}_a {i(\cancel{k}+m_B)\over k^2-m_B^2} g_{MB}^N \cancel{p}_M \gamma_5) u_N\;,
\end{eqnarray}
where $p_M$ denotes the outgoing meson momentum in final states, $k$ denotes the baryon momentum in propagator.
The spin-averaged squared matrix element of $B\to \ell a$ with $|\Delta(B-L)|=2$ reads as
\begin{eqnarray}
\overline{|\mathcal{M}|^2}&=&(|m_{Br,L}|^2+|m_{Br,R}|^2)(2p_{\ell_r}\cdot p_a p_B\cdot p_a-p_{\ell_r}\cdot p_B m_a^2)+2m_B m_a^2 m_{\ell_r} {\rm Re}(m_{Br,L}m_{Br,R}^\ast)\nonumber \\
&=&{1\over 2}(|m_{Br,L}|^2+|m_{Br,R}|^2)[(m_B^2-m_{\ell_r}^2)^2-m_a^2(m_B^2+m_{\ell_r}^2)]+2m_B m_a^2 m_{\ell_r} {\rm Re}(m_{Br,L}m_{Br,R}^\ast)\;,\nonumber \\
\end{eqnarray}
The matrix element of $B\to \bar{\ell} a$ with $|\Delta(B-L)|=0$ turns out to be the same as above.
It is obtained by substituting $\ell\to \ell^c$, $P_{\bar{X}}\to P_X$ in the Lagrangian and $P_{\bar{X}}\to P_X$ in the matrix elements.
For brevity, we collect the spin-averaged squared matrix elements of three-body BNV nucleon decays $N\to M \ell a$ and $N\to M \bar{\ell} a$ in App.~\ref{app:3body}.

The decays $n\to \pi^+\ell^- a$ and $n\to \bar K^0 \nu a$ with $|\Delta(B-L)|=2$ and the $|\Delta(B-L)|=0$ decays
$n\to K^- \ell^+ a$, $p\to \bar K^0 \ell^+ a$, and $n\to \bar K^0 \bar{\nu} a$
are not induced by dimension-8 aLEFT operators at leading order of chiral power counting due to their isospin $I$ and strangeness $S$ properties.
For the $|\Delta (B-L)|=2$ decays $n\to \pi^+\ell^-a$ and $n\to \bar K^0 \nu a$, the quantum numbers are $(\Delta I,\Delta S)=(\tfrac32,0)$ and $(\Delta I,\Delta S)=(1,-1)$, respectively.
Similarly, the $|\Delta(B-L)|=0$ decays $n\to K^-\ell^+a$ and $p\to \bar K^0\ell^+ a$ have $(\Delta I,\Delta S)=(0,-1)$ and $n\to \bar K^0 \bar \nu a$ has $(\Delta I,\Delta S)=(1,-1)$. The absence of those decays in Tabs.~\ref{tab:3bodyBL2} and \ref{tab:3bodyBL0} are indicated with dashes.
The maximum isospin change $\Delta I$ is $1$ for the operators in Tabs.~\ref{tab:aLEFTBL2} and \ref{tab:aLEFTBL0} which contribute at leading order in chiral power counting. Isospin $|\Delta I|=\frac32$ requires three identical quarks which is not allowed due to the antisymmetry of the scalar bilinear. Although there are operators with $(\Delta I,\Delta S) = (1,-1)$, namely $[\mathcal{O}_{\partial a eddd}^{VX,SX}]_{r223}$ with $X=L,R$, they consist of three down-type quarks and thus charge conservation requires the fourth fermion to be a charged lepton. Finally, although there are operators with $(\Delta I,\Delta S) =(0,-1)$, like $[\mathcal{O}^{VL,SL}_{\partial ad\nu ud}]_{2r13}$ and $[\mathcal{O}^{VL,SR}_{\partial ad\nu du}]_{2r31}$, electric charge conseration requires the combination of an up-, down- and strange quark to be accompanied by a neutrino.

\begin{table}[htbp!]
\centering
\renewcommand{\arraystretch}{1.2}
\begin{tabular}{c|c}
\hline
Process & $m_{Br,X}$ \\
\hline
$n\to \nu a$ & $m_{nr,L}=[C^{VL,SL}_{\partial a \nu ddu}]_{r221} (-\beta)$ \\
\hline
$\Lambda^0\to \nu a$ & $m_{\Lambda r,L}=[C^{VL,SL}_{\partial a \nu ddu}]_{r231} ({1\over \sqrt{6}}\beta)+[C^{VL,SL}_{\partial a \nu ddu}]_{r321} (\sqrt{2\over 3}\beta)$ \\
\hline
$\Sigma^0\to \nu a$ & $m_{\Sigma r,L}=[C^{VL,SL}_{\partial a \nu ddu}]_{r231} (-{1\over \sqrt{2}}\beta)$ \\
\hline
$\Xi^0\to \nu a$ & $m_{\Xi r,L}=[C^{VL,SL}_{\partial a \nu ddu}]_{r331} (\beta)$ \\
\hline
$\Sigma^-\to \ell^- a$ & $m_{\Sigma r,L}=[C^{VL,SL}_{\partial a eddd}]_{r223} (\beta)$ \\
 & $m_{\Sigma r,R}=[C^{VR,SR}_{\partial a eddd}]_{r223} (-\beta)$ \\
\hline
$\Xi^-\to \ell^- a$ & $m_{\Xi r,L}=[C^{VL,SL}_{\partial a eddd}]_{r323} (\beta)$ \\
 & $m_{\Xi r,R}=[C^{VR,SR}_{\partial a eddd}]_{r323} (-\beta)$ \\
\hline
\end{tabular}
\caption{
The couplings for two-body BNV decays $B\to \ell a$ with $|\Delta (B-L)|=2$.
}
\label{tab:2bodyBL2}
\end{table}

\begin{table}[htbp!]
\centering
\renewcommand{\arraystretch}{1.2}
\begin{tabular}{c|c}
\hline
Process & $m_{Br,X}$ \\
\hline
$p\to \ell^+ a$ & $m_{pr,L}=[C^{VR,SR}_{\partial a euud}]_{r112} (-\beta)$ \\
 & $m_{pr,R}=[C^{VL,SR}_{\partial a eudu}]_{r121} (-\alpha)+[C^{VL,SL}_{\partial a eudu}]_{r121} (-\beta)$ \\
\hline
$\Sigma^+\to \ell^+ a$ & $m_{\Sigma r,L}=[C^{VR,SR}_{\partial a euud}]_{r113} (\beta)$ \\
 & $m_{\Sigma r,R}=[C^{VL,SR}_{\partial a eudu}]_{r131} (\alpha)+[C^{VL,SL}_{\partial a eudu}]_{r131} (\beta)$ \\
\hline
$n\to \bar{\nu}a$ & $m_{nr,L}=[C^{VL,SL}_{\partial a d\nu ud}]_{2r12} (\alpha)+[C^{VL,SR}_{\partial a d\nu du}]_{2r21} (-\beta)$\\
\hline
$\Lambda^0\to \bar{\nu}a$ & $m_{\Lambda r,L}=[C^{VL,SL}_{\partial a d\nu ud}]_{2r13} (-{1\over \sqrt{6}}\alpha)+[C^{VL,SL}_{\partial a d\nu ud}]_{3r12} (-\sqrt{2\over 3}\alpha)$\\
& $+[C^{VL,SR}_{\partial a d\nu du}]_{2r31} ({1\over \sqrt{6}}\beta)+[C^{VL,SR}_{\partial a d\nu du}]_{3r21} (\sqrt{2\over 3}\beta)$\\
\hline
$\Sigma^0\to \bar{\nu}a$ & $m_{\Sigma r,L}=[C^{VL,SL}_{\partial a d\nu ud}]_{2r13} ({1\over \sqrt{2}}\alpha)+[C^{VL,SR}_{\partial a d\nu du}]_{2r31} (-{1\over \sqrt{2}}\beta)$\\
\hline
$\Xi^0\to \bar{\nu}a$ & $m_{\Xi r,L}=[C^{VL,SL}_{\partial a d\nu ud}]_{3r13} (-\alpha)+[C^{VL,SR}_{\partial a d\nu du}]_{3r31} (\beta)$\\
\hline
\end{tabular}
\caption{
The couplings for two-body BNV decays $B\to \bar{\ell}a$ with $|\Delta (B-L)|=0$.
}
\label{tab:2bodyBL0}
\end{table}

\begin{table}[htbp!]
\centering
\renewcommand{\arraystretch}{1.2}
\hspace*{-0.05\columnwidth}\resizebox{1.1\columnwidth}{!}{
\begin{tabular}{c|c|c|c}
\hline
Process & $g_{MB}^N$ & $m_{Br,X}$ & $y_{Mr,X}^N$\\
\hline
$p\to K^+ \nu a$ & $g_{K\Sigma}^p={D-F\over \sqrt{2}f_\pi}$ & $m_{\Sigma r,L}=[C^{VL,SL}_{\partial a \nu ddu}]_{r231} (-\beta{1\over \sqrt{2}})$ & $y_{Kr,L}^p=[C^{VL,SL}_{\partial a \nu ddu}]_{r321}(-{\beta\over f_\pi})$\\
 & $g_{K\Lambda}^p=-{D+3F\over \sqrt{6}f_\pi}$ & $m_{\Lambda r,L}=[C^{VL,SL}_{\partial a \nu ddu}]_{r231} (\beta{1\over \sqrt{6}})+[C^{VL,SL}_{\partial a \nu ddu}]_{r321} (\beta \sqrt{2\over 3})$ & \\
\hline
$n\to \pi^0 \nu a$ & $g_{\pi n}^n=-{D+F\over \sqrt{2}f_\pi}$ & $m_{n r,L}=[C^{VL,SL}_{\partial a \nu ddu}]_{r221} (-\beta)$ & $y_{\pi r,L}^n=[C^{VL,SL}_{\partial a \nu ddu}]_{r221}({\beta\over \sqrt{2}f_\pi})$\\
\hline
$p\to \pi^+ \nu a$ & $g_{\pi n}^p={D+F\over f_\pi}$ & $m_{n r,L}=[C^{VL,SL}_{\partial a \nu ddu}]_{r221} (-\beta)$ & $y_{\pi r,L}^p=[C^{VL,SL}_{\partial a \nu ddu}]_{r221}(-{\beta\over f_\pi})$ \\
\hline
$n\to \eta^0 \nu a$ & $g_{\eta n}^n={3F-D\over \sqrt{6}f_\pi}$ & $m_{n r,L}=[C^{VL,SL}_{\partial a \nu ddu}]_{r221} (-\beta)$ & $y_{\eta r,L}^n=[C^{VL,SL}_{\partial a \nu ddu}]_{r221}(-\sqrt{3\over 2}{\beta\over f_\pi})$ \\
\hline
$n\to K^+ \ell^- a$ & $g_{K\Sigma}^n={D-F\over f_\pi}$ & $m_{\Sigma r,L}=[C^{VL,SL}_{\partial a e ddd}]_{r223} (\beta)$ & $y_{K r,L}^n=[C^{VL,SL}_{\partial a e ddd}]_{r223}(-{\beta\over f_\pi})$ \\
 &  & $m_{\Sigma r,R}=[C^{VR,SR}_{\partial a e ddd}]_{r223} (-\beta)$ & $y_{K r,R}^n=[C^{VR,SR}_{\partial a e ddd}]_{r223}(-{\beta\over f_\pi})$ \\
\hline
$n\to K^0 \nu a$ & $g_{K\Sigma}^n=-{D-F\over \sqrt{2}f_\pi}$ & $m_{\Sigma r,L}=[C^{VL,SL}_{\partial a \nu ddu}]_{r231} (-\beta{1\over \sqrt{2}})$ & $y_{Kr,L}^n=[C^{VL,SL}_{\partial a \nu ddu}]_{r231}(-{\beta\over f_\pi})+[C^{VL,SL}_{\partial a \nu ddu}]_{r321}(-{\beta\over f_\pi})$ \\
 & $g_{K\Lambda}^n=-{D+3F\over \sqrt{6}f_\pi}$ & $m_{\Lambda r,L}=[C^{VL,SL}_{\partial a \nu ddu}]_{r231} (\beta{1\over \sqrt{6}})+[C^{VL,SL}_{\partial a \nu ddu}]_{r321} (\beta \sqrt{2\over 3})$ & \\
\hline
$n\to \pi^+ \ell^- a$ & $-$ & $-$ & $-$\\
\hline
$n\to \bar{K}^0 \nu a$ & $-$ & $-$ & $-$ \\
\hline
\end{tabular}
}
\caption{
The couplings for three-body BNV decays $N\to M\ell a$ with $|\Delta (B-L)|=2$.
}
\label{tab:3bodyBL2}
\end{table}

\begin{table}[htbp!]
\renewcommand{\arraystretch}{1.2}
\hspace*{-0.05\columnwidth}
\resizebox{1.1\columnwidth}{!}{
\begin{tabular}{c|c|c|c}
\hline
Process & $g_{MB}^N$ & $m_{Br,X}$ & $y_{Mr,X}^N$\\
\hline
$p\to \pi^0 \ell^+ a$ & $g_{\pi p}^p={D+F\over \sqrt{2}f_\pi}$ & $m_{pr,L}=[C^{VR,SR}_{\partial a euud}]_{r112} (-\beta)$ & $y_{\pi r,L}^p=[C^{VR,SR}_{\partial a euud}]_{r112}({1\over \sqrt{2}}{\beta\over f_\pi})$\\
&  & $m_{pr,R}=[C^{VL,SR}_{\partial a eudu}]_{r121} (-\alpha)+[C^{VL,SL}_{\partial a eudu}]_{r121} (-\beta)$ & $y_{\pi r,R}^p=[C^{VL,SR}_{\partial a eudu}]_{r121}(-{1\over \sqrt{2}}{\alpha\over f_\pi})$\\
& & & $+[C^{VL,SL}_{\partial a eudu}]_{r121}(-{1\over \sqrt{2}}{\beta\over f_\pi})$\\
\hline
$p\to \eta^0 \ell^+ a$ & $g_{\eta p}^p={3F-D\over \sqrt{6}f_\pi}$ & $m_{pr,L}=[C^{VR,SR}_{\partial a euud}]_{r112} (-\beta)$ & $y_{\eta r,L}^p=[C^{VR,SR}_{\partial a euud}]_{r112}(\sqrt{3\over 2}{\beta\over f_\pi})$\\
&  & $m_{pr,R}=[C^{VL,SR}_{\partial a eudu}]_{r121} (-\alpha)+[C^{VL,SL}_{\partial a eudu}]_{r121} (-\beta)$ & $y_{\eta r,R}^p=[C^{VL,SR}_{\partial a eudu}]_{r121}({1\over \sqrt{6}}{\alpha\over f_\pi})$\\
& & & $+[C^{VL,SL}_{\partial a eudu}]_{r121}(-\sqrt{3\over 2}{\beta\over f_\pi})$\\
\hline
$p\to K^+ \bar{\nu} a$ & $g_{K\Sigma}^p={D-F\over \sqrt{2}f_\pi}$ & $m_{\Sigma r,L}=[C^{VL,SL}_{\partial a d\nu ud}]_{2r13} ({1\over \sqrt{2}}\alpha)+[C^{VL,SR}_{\partial a d\nu du}]_{2r31} (-{1\over \sqrt{2}}\beta)$ & $y_{Kr,L}^p=[C^{VL,SL}_{\partial a d\nu ud}]_{3r12}(-{\alpha\over f_\pi})+[C^{VL,SR}_{\partial a d\nu du}]_{3r21}({\beta\over f_\pi})$\\
& $g_{K\Lambda}^p=-{D+3F\over \sqrt{6}f_\pi}$ & $m_{\Lambda r,L}=[C^{VL,SL}_{\partial a d\nu ud}]_{2r13} (-{1\over \sqrt{6}}\alpha)+[C^{VL,SL}_{\partial a d\nu ud}]_{3r12} (-\sqrt{2\over 3}\alpha)$ & \\
& & $+[C^{VL,SR}_{\partial a d\nu du}]_{2r31} ({1\over \sqrt{6}}\beta)+[C^{VL,SR}_{\partial a d\nu du}]_{3r21} (\sqrt{2\over 3}\beta)$ & \\
\hline
$n\to \pi^- \ell^+ a$ & $g_{\pi p}^n={D+F\over f_\pi}$ & $m_{pr,L}=[C^{VR,SR}_{\partial a euud}]_{r112} (-\beta)$ & $y_{\pi r,L}^n=[C^{VR,SR}_{\partial a euud}]_{r112}({\beta\over f_\pi})$ \\
 & & $m_{pr,R}=[C^{VL,SR}_{\partial a eudu}]_{r121} (-\alpha)+[C^{VL,SL}_{\partial a eudu}]_{r121} (-\beta)$ & $y_{\pi r,R}^n=[C^{VL,SR}_{\partial a eudu}]_{r121}(-{\alpha\over f_\pi})+[C^{VL,SL}_{\partial a eudu}]_{r121}(-{\beta\over f_\pi})$\\
\hline
$n\to \pi^0 \bar{\nu} a$ & $g_{\pi n}^n=-{D+F\over \sqrt{2}f_\pi}$ & $m_{nr,L}=[C^{VL,SL}_{\partial a d\nu ud}]_{2r12} (\alpha)+[C^{VL,SR}_{\partial a d\nu du}]_{2r21} (-\beta)$ & $y_{\pi r,L}^n=[C^{VL,SL}_{\partial a d\nu ud}]_{2r12}({\alpha\over \sqrt{2}f_\pi})$\\
& & & $+[C^{VL,SR}_{\partial a d\nu du}]_{2r21}(-{\beta\over \sqrt{2}f_\pi})$\\
\hline
$p\to K^0 \ell^+ a$ & $g_{K\Sigma}^p={D-F\over f_\pi}$ & $m_{\Sigma r,L}=[C^{VR,SR}_{\partial a euud}]_{r113} (\beta)$ & $y_{Kr,L}^p=[C^{VR,SR}_{\partial a euud}]_{r113} ({\beta\over f_\pi}) $\\
& & $m_{\Sigma r,R}=[C^{VL,SR}_{\partial a eudu}]_{r131} (\alpha)+[C^{VL,SL}_{\partial a eudu}]_{r131} (\beta)$ & $y_{K r,R}^p=[C^{VL,SR}_{\partial a eudu}]_{r131}({\alpha\over f_\pi})+[C^{VL,SL}_{\partial a eudu}]_{r131}(-{\beta\over f_\pi})$\\
\hline
$p\to \pi^+ \bar{\nu} a$ & $g_{\pi n}^p={D+F\over f_\pi}$ & $m_{nr,L}=[C^{VL,SL}_{\partial a d\nu ud}]_{2r12} (\alpha)+[C^{VL,SR}_{\partial a d\nu du}]_{2r21} (-\beta)$ & $y_{\pi r,L}^p=[C^{VL,SL}_{\partial a d\nu ud}]_{2r12}(-{\alpha\over f_\pi})+[C^{VL,SR}_{\partial a d\nu du}]_{2r21}({\beta\over f_\pi})$\\
\hline
$n\to \eta^0 \bar{\nu} a$ & $g_{\eta n}^n={3F-D\over \sqrt{6}f_\pi}$ & $m_{nr,L}=[C^{VL,SL}_{\partial a d\nu ud}]_{2r12} (\alpha)+[C^{VL,SR}_{\partial a d\nu du}]_{2r21} (-\beta)$ & $y_{\eta r, L}^n=[C^{VL,SL}_{\partial a d\nu ud}]_{2r12}({1\over \sqrt{6}}{\alpha\over f_\pi})$\\
& & & $+[C^{VL,SR}_{\partial a d\nu du}]_{2r21}(\sqrt{3\over 2}{\beta\over f_\pi})$\\
\hline
$n\to K^0 \bar{\nu} a$ & $g_{K \Sigma}^n=-{D-F\over \sqrt{2}f_\pi}$ & $m_{\Sigma r,L}=[C^{VL,SL}_{\partial a d\nu ud}]_{2r13} ({1\over \sqrt{2}}\alpha)+[C^{VL,SR}_{\partial a d\nu du}]_{2r31} (-{1\over \sqrt{2}}\beta)$ & $y_{Kr,L}^n=[C^{VL,SL}_{\partial a d\nu ud}]_{2r13}({\alpha\over f_\pi})+[C^{VL,SL}_{\partial a d\nu ud}]_{3r12}(-{\alpha\over f_\pi})$\\
 & $g_{K \Lambda}^n=-{D+3F\over \sqrt{6}f_\pi}$ & $m_{\Lambda r,L}=[C^{VL,SL}_{\partial a d\nu ud}]_{2r13} (-{1\over \sqrt{6}}\alpha)+[C^{VL,SL}_{\partial a d\nu ud}]_{3r12} (-\sqrt{2\over 3}\alpha)$ & $+[C^{VL,SR}_{\partial a d\nu du}]_{2r31}({\beta\over f_\pi})+[C^{VL,SR}_{\partial a d\nu du}]_{3r21}({\beta\over f_\pi})$\\
& & $+[C^{VL,SR}_{\partial a d\nu du}]_{2r31} ({1\over \sqrt{6}}\beta)+[C^{VL,SR}_{\partial a d\nu du}]_{3r21} (\sqrt{2\over 3}\beta)$ & \\
\hline
$n\to K^- \ell^+ a$ & $-$ & $-$ & $-$ \\
\hline
$p\to \bar{K}^0 \ell^+ a$ & $-$ & $-$ & $-$ \\
\hline
$n\to \bar{K}^0 \bar{\nu} a$ & $-$ & $-$ & $-$ \\
\hline
\end{tabular}
}
\caption{
The couplings for three-body BNV decays $N\to M\bar{\ell} a$ with $|\Delta (B-L)|=0$.
}
\label{tab:3bodyBL0}
\end{table}

\section{Numerical results}
\label{sec:Results}

In this section, we show the numerical results of the two-body and three-body BNV decay rates.
We adopt the following hadronic parameters
for numerical calculations
\begin{eqnarray}
&&\alpha = -0.01257(111)~{\rm GeV}^3~\text{\cite{Yoo:2021gql}}\;,~~~\beta= 0.01269(107)~{\rm GeV}^3~\text{\cite{Yoo:2021gql}}\;, \\
&&D=0.730(11)~\text{\cite{Bali:2022qja}}\;,~~~F=0.447^{(6)}_{(7)}~\text{\cite{Bali:2022qja}}\;,~~~f_\pi= 130.41(20)~{\rm MeV}~\text{\cite{Workman:2022ynf}}\;.
\end{eqnarray}
The ALP mass $m_a$ is approximately set as zero in the numerical calculation. In the following subsections, we present our main results.

\subsection{Two-body BNV baryon decays $B\to \ell a$}
\label{sec:Bla}

The numerical results of
nucleon partial decay widths and baryon decay branching fractions
are as follows.~\footnote{For the decays with SM neutrino in final states here and below, we use ``$r$'' to denote the undetermined neutrino flavor.}
Here we only show the formulae for $|\Delta (B-L)|=2$ and those for $|\Delta (B-L)|=0$ are collected in App.~\ref{app:BL0}.


(1) $n\to \nu a$
\begin{eqnarray}
\Gamma(n\to \nu a)
={|[C^{VL,SL}_{\partial a \nu ddu}]_{r221}|^2 \over 1.6\times 10^{-59}~{\rm GeV}^{-8}} {1\over 10^{33}~{\rm years}}  \;,
\end{eqnarray}

(2) $\Lambda^0\to \nu a$
\begin{eqnarray}
{\rm BR}(\Lambda^0\to \nu a)&=&1.5\times 10^8~{\rm GeV}^8 |[C^{VL,SL}_{\partial a \nu ddu}]_{r231}|^2 + 5.9\times 10^8~{\rm GeV}^9 |[C^{VL,SL}_{\partial a \nu ddu}]_{r321}|^2 \nonumber \\
&&+5.9\times 10^8~{\rm GeV}^8 {\rm Re}([C^{VL,SL}_{\partial a \nu ddu}]_{r231} [C^{VL,SL}_{\partial a \nu ddu}]_{r321}^\ast)\;,
\end{eqnarray}

(3) $\Sigma^0\to \nu a$
\begin{eqnarray}
{\rm BR}(\Sigma^0\to \nu a)=0.15~{\rm GeV}^8 |[C^{VL,SL}_{\partial a \nu ddu}]_{r231}|^2 \;,
\end{eqnarray}

(4) $\Xi^0\to \nu a$
\begin{eqnarray}
{\rm BR}(\Xi^0\to \nu a)=1.6\times 10^9~{\rm GeV}^8 |[C^{VL,SL}_{\partial a \nu ddu}]_{r331}|^2 \;,
\end{eqnarray}

(5) $\Sigma^-\to e^-(\mu^-) a$
\begin{eqnarray}
{\rm BR}(\Sigma^-\to e^- a)=6.2\times 10^8~{\rm GeV}^8 (|[C^{VL,SL}_{\partial a eddd}]_{1223}|^2 + |[C^{VR,SR}_{\partial a eddd}]_{1223}|^2)\;,
\end{eqnarray}

\begin{eqnarray}
{\rm BR}(\Sigma^-\to \mu^- a)=6.0\times 10^8~{\rm GeV}^8 (|[C^{VL,SL}_{\partial a eddd}]_{2223}|^2 + |[C^{VR,SR}_{\partial a eddd}]_{2223}|^2)\;,
\end{eqnarray}

(6) $\Xi^-\to e^-(\mu^-) a$
\begin{eqnarray}
{\rm BR}(\Xi^-\to e^- a)=9.2\times 10^8~{\rm GeV}^8 (|[C^{VL,SL}_{\partial a eddd}]_{1323}|^2 + |[C^{VR,SR}_{\partial a eddd}]_{1323}|^2)\;,
\end{eqnarray}

\begin{eqnarray}
{\rm BR}(\Xi^-\to \mu^- a)=9.2\times 10^8~{\rm GeV}^8 (|[C^{VL,SL}_{\partial a eddd}]_{2323}|^2 + |[C^{VR,SR}_{\partial a eddd}]_{2323}|^2)\;,
\end{eqnarray}

\subsection{Three-body BNV nucleon decays $N\to M \ell a$}
\label{sec:NMla}

The numerical results of nucleon partial decay widths with $|\Delta (B-L)|=2$ are as follows. Those with $|\Delta (B-L)|=0$ are also shown in App.~\ref{app:BL0}.

(1) $p\to K^+ \nu a$

\begin{eqnarray}
\Gamma(p\to K^+ \nu a)
&=&\Big({|[C^{VL,SL}_{\partial a \nu ddu}]_{r231}|^2\over 1.5\times 10^{-55}~{\rm GeV}^{-8}}+{|[C^{VL,SL}_{\partial a \nu ddu}]_{r321}|^2\over 6.4\times 10^{-57}~{\rm GeV}^{-8}}\nonumber\\
&+&{{\rm Re}([C^{VL,SL}_{\partial a \nu ddu}]_{r231}[C^{VL,SL}_{\partial a \nu ddu}]_{r321}^*)\over 1.6\times 10^{-56}~{\rm GeV}^{-8}}\Big){1\over 10^{33}~{\rm years}}\;,
\end{eqnarray}

(2) $n\to \pi^0 \nu a$

\begin{eqnarray}
\Gamma(n\to \pi^0 \nu a)
={|[C^{VL,SL}_{\partial a \nu ddu}]_{r221}|^2\over 4.0\times 10^{-58}~{\rm GeV}^{-8}}{1\over 10^{33}~{\rm years}}\;,
\end{eqnarray}

(3) $p\to \pi^+ \nu a$

\begin{eqnarray}
\Gamma(p\to \pi^+ \nu a)
={|[C^{VL,SL}_{\partial a \nu ddu}]_{r221}|^2\over 2.1\times 10^{-58}~{\rm GeV}^{-8}}{1\over 10^{33}~{\rm years}}\;,
\end{eqnarray}

(4) $n\to \eta^0 \nu a$

\begin{eqnarray}
\Gamma(n\to \eta^0 \nu a)
={|[C^{VL,SL}_{\partial a \nu ddu}]_{r221}|^2\over 1.3\times 10^{-56}~{\rm GeV}^{-8}}{1\over 10^{33}~{\rm years}}\;,
\end{eqnarray}

(5) $n\to K^+ e^-(\mu^-) a$

\begin{eqnarray}
\Gamma(n\to K^+ e^- a)
&=&{|[C^{VL,SL}_{\partial a eddd}]_{1223}|^2+|[C^{VR,SR}_{\partial a eddd}]_{1223}|^2\over 1.6\times 10^{-56}~{\rm GeV}^{-8}}{1\over 10^{33}~{\rm years}}\;,
\end{eqnarray}

\begin{eqnarray}
\Gamma(n\to K^+ \mu^- a)
&=&{|[C^{VL,SL}_{\partial a eddd}]_{2223}|^2+|[C^{VR,SR}_{\partial a eddd}]_{2223}|^2\over 2.2\times 10^{-56}~{\rm GeV}^{-8}}{1\over 10^{33}~{\rm years}}\;,
\end{eqnarray}

(6) $n\to K^0 \nu a$

\begin{eqnarray}
\Gamma(n\to K^0 \nu a)
&=&\Big({|[C^{VL,SL}_{\partial a \nu ddu}]_{r231}|^2\over 1.0\times 10^{-56}~{\rm GeV}^{-8}}+{|[C^{VL,SL}_{\partial a \nu ddu}]_{r321}|^2 \over 6.6\times 10^{-57}~{\rm GeV}^{-8}}\nonumber\\
&+&{{\rm Re}([C^{VL,SL}_{\partial a \nu ddu}]_{r231}[C^{VL,SL}_{\partial a \nu ddu}]_{r321}^*)\over 4.1\times 10^{-57}~{\rm GeV}^{-8}}\Big){1\over 10^{33}~{\rm years}}\;,
\end{eqnarray}

\begin{figure}[h!]
\centering
\includegraphics[width=0.36\textwidth]{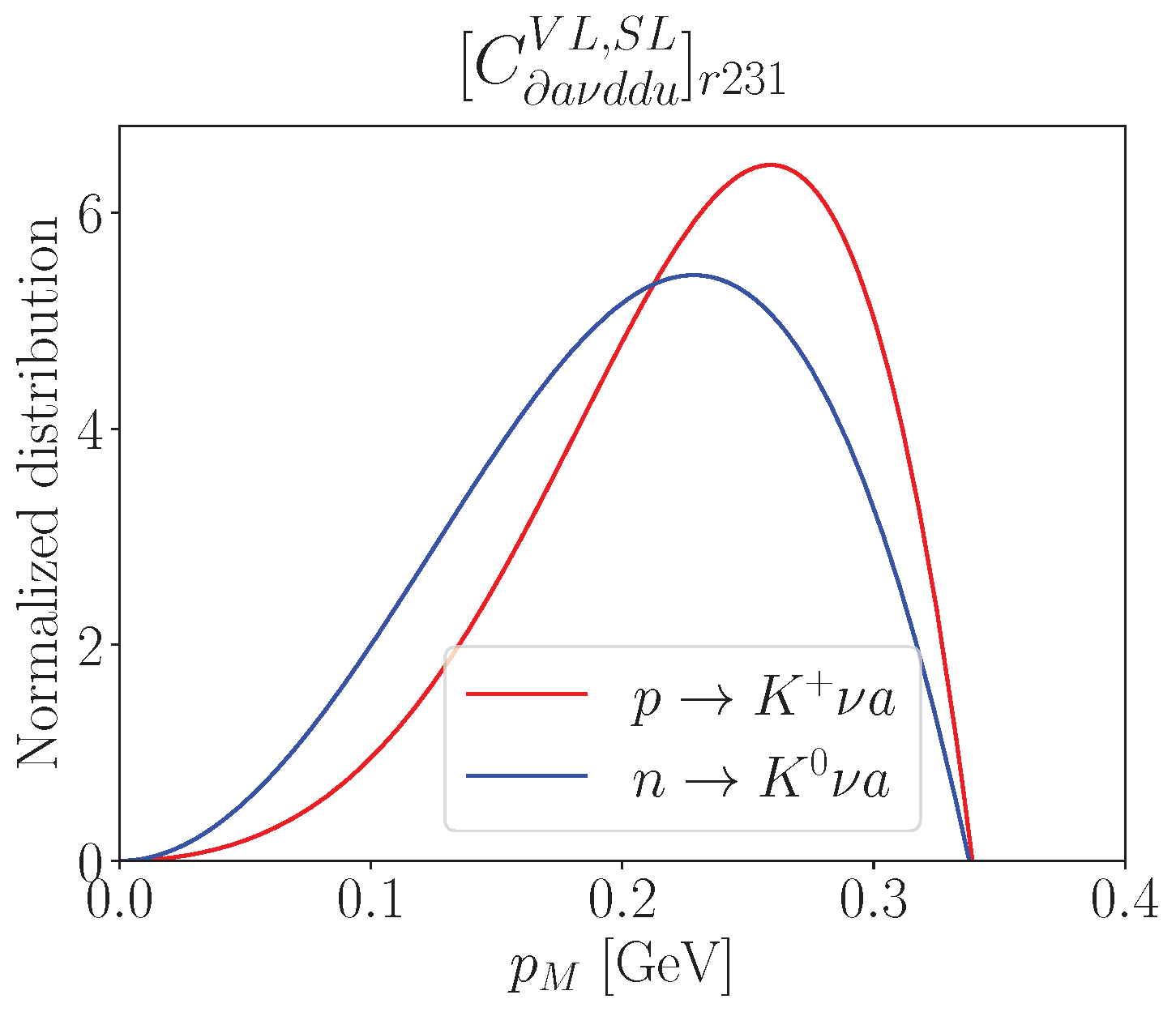}
\includegraphics[width=0.36\textwidth]{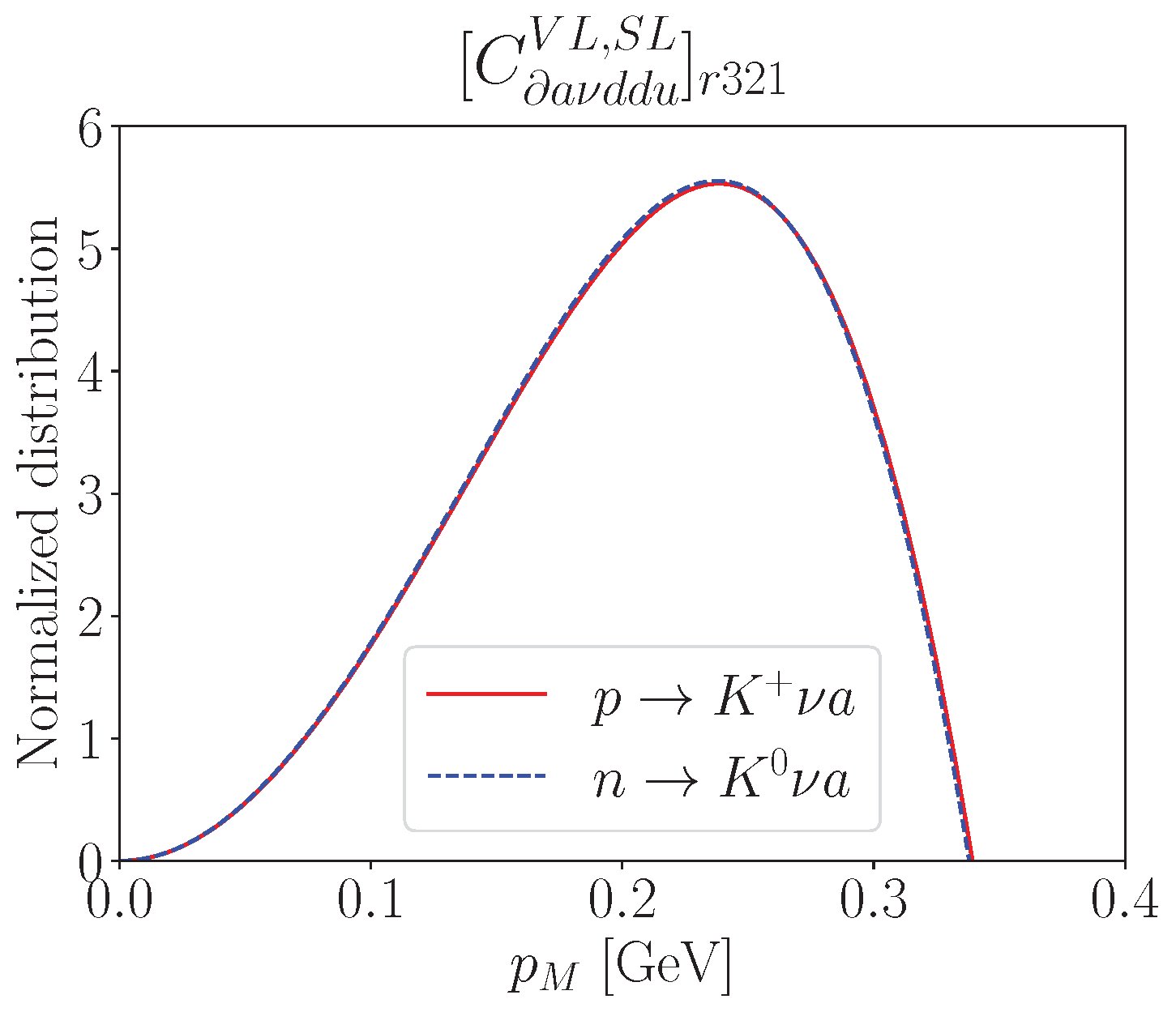}
\includegraphics[width=0.36\textwidth]{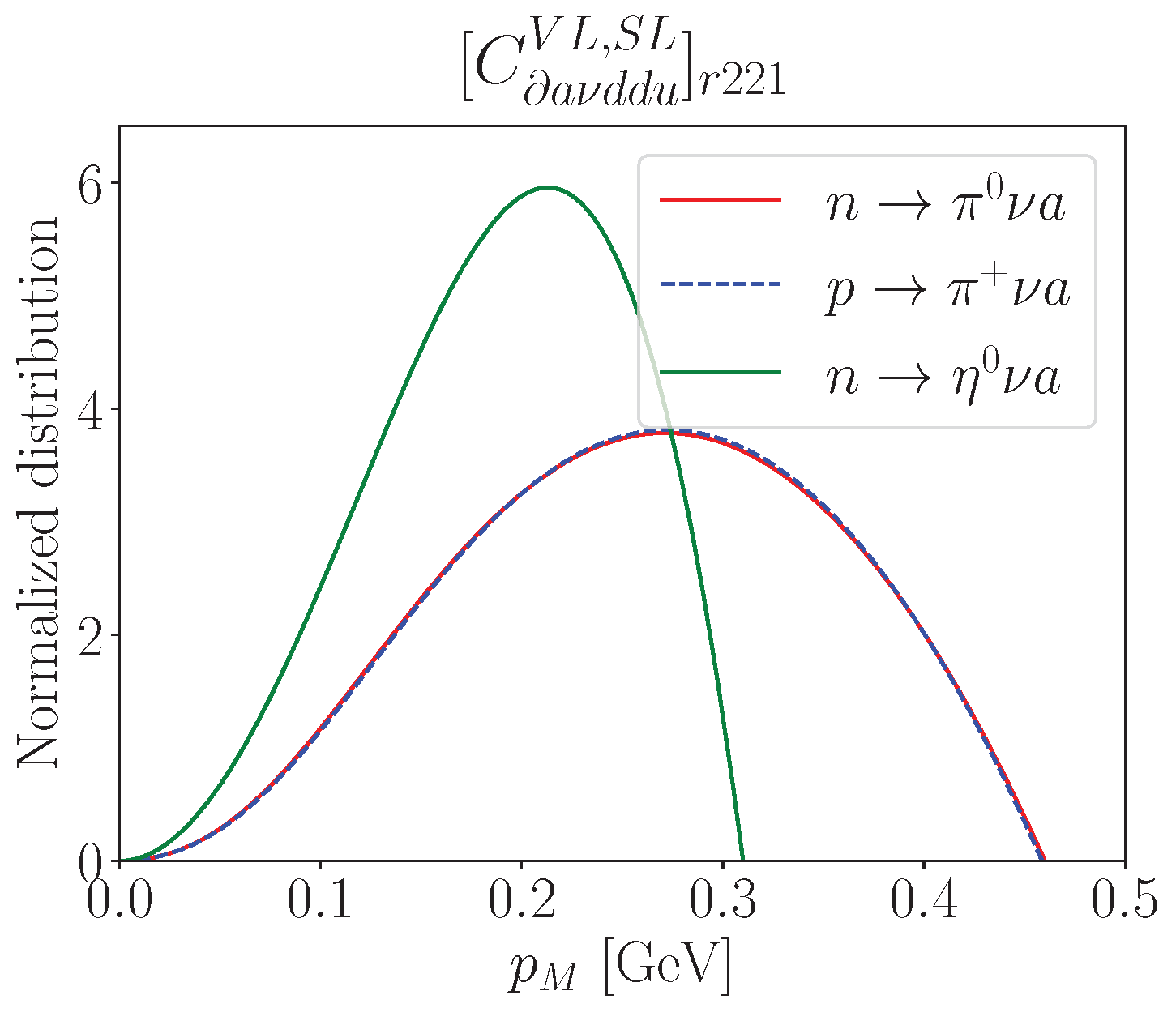}
\includegraphics[width=0.36\textwidth]{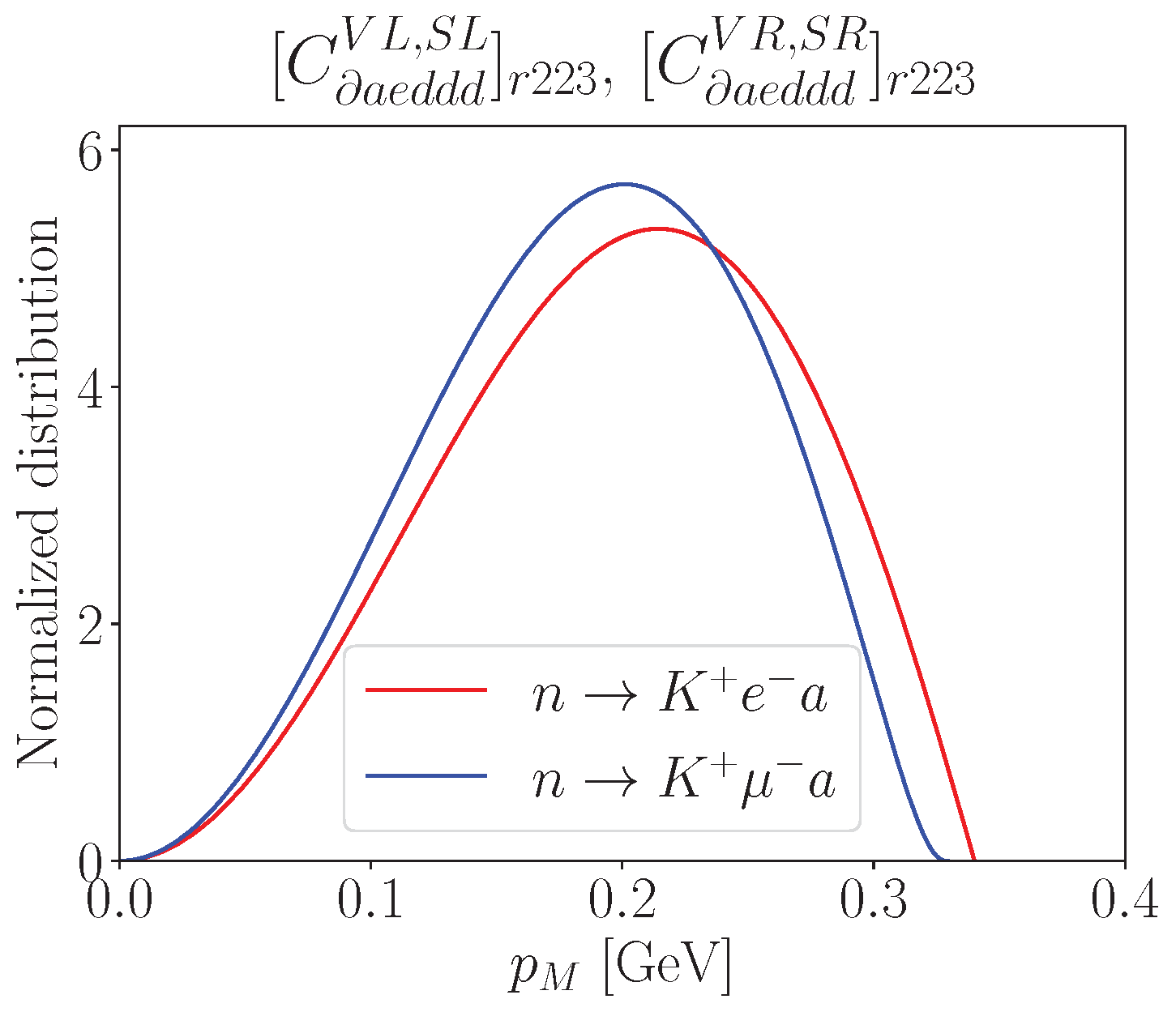}
\caption{The normalized momentum distribution of the final state meson in the rest frame of the nucleon for the processes $N\to M \ell a$ with $|\Delta(B-L)|=2$. The details are described in the text.
}
\label{fig:PM_BL2}
\end{figure}

\begin{figure}[h!]
\centering
\minigraph{4.5cm}{-0.05in}{}{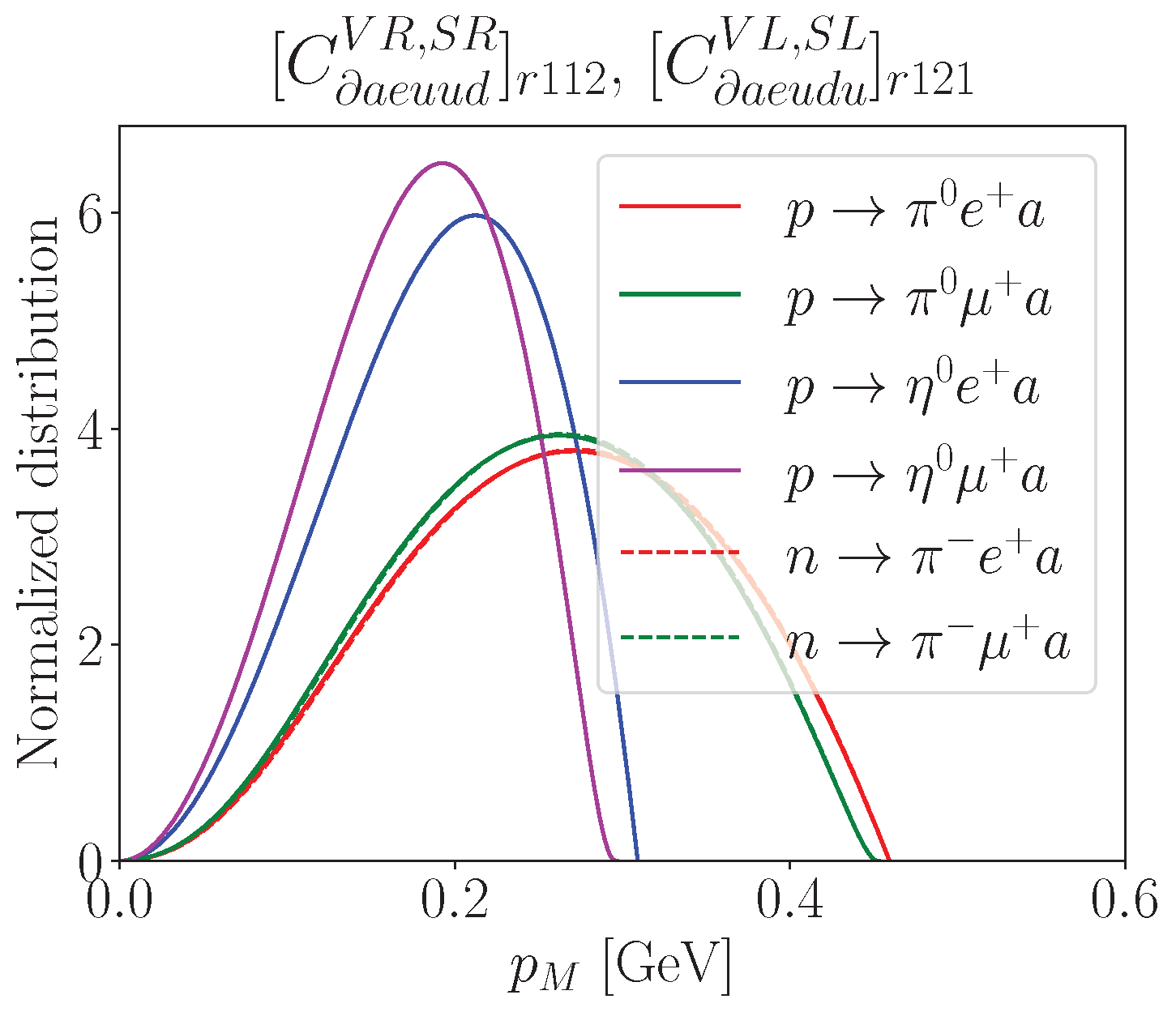}
\minigraph{4.5cm}{-0.05in}{}{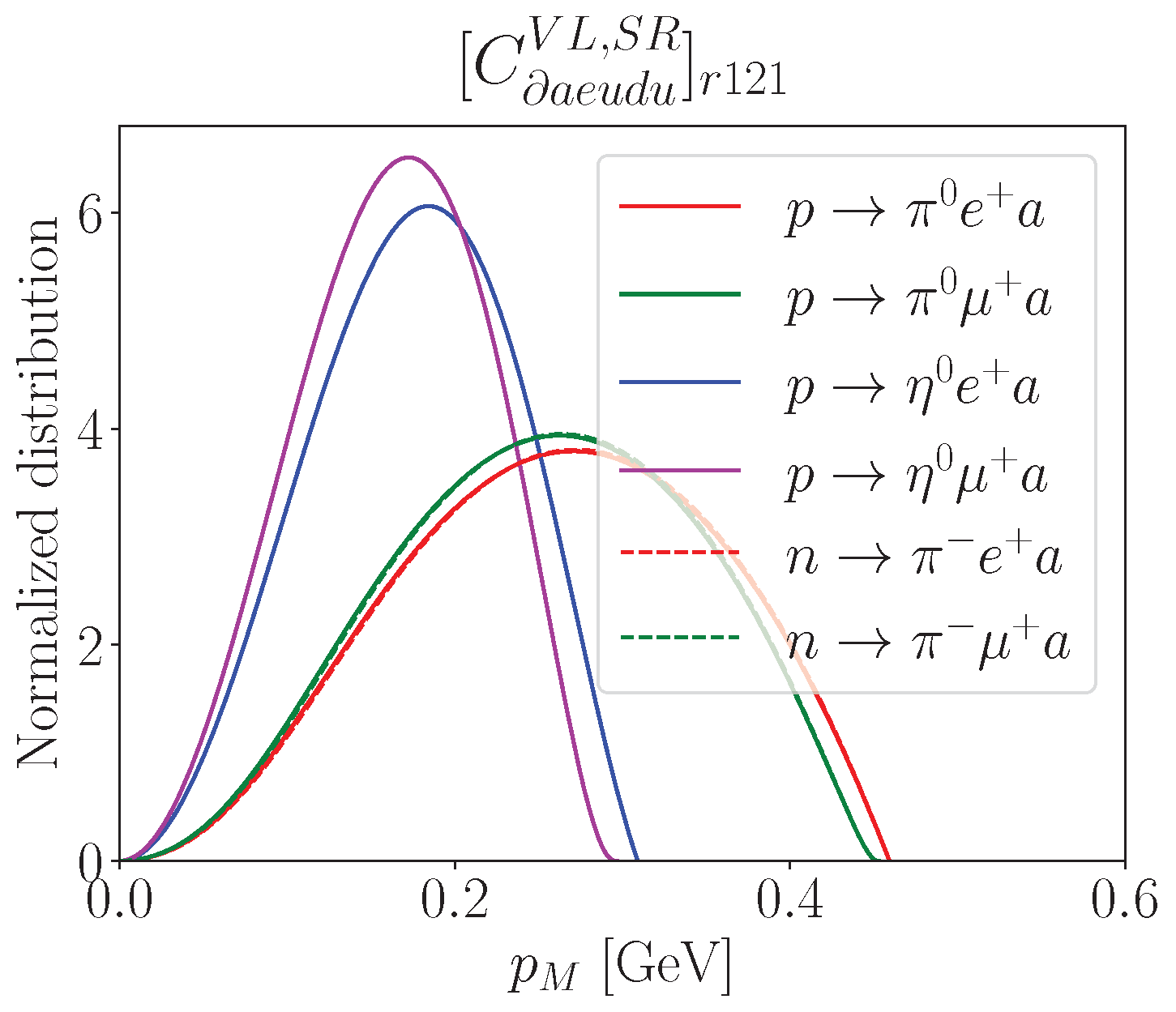}
\minigraph{4.5cm}{-0.05in}{}{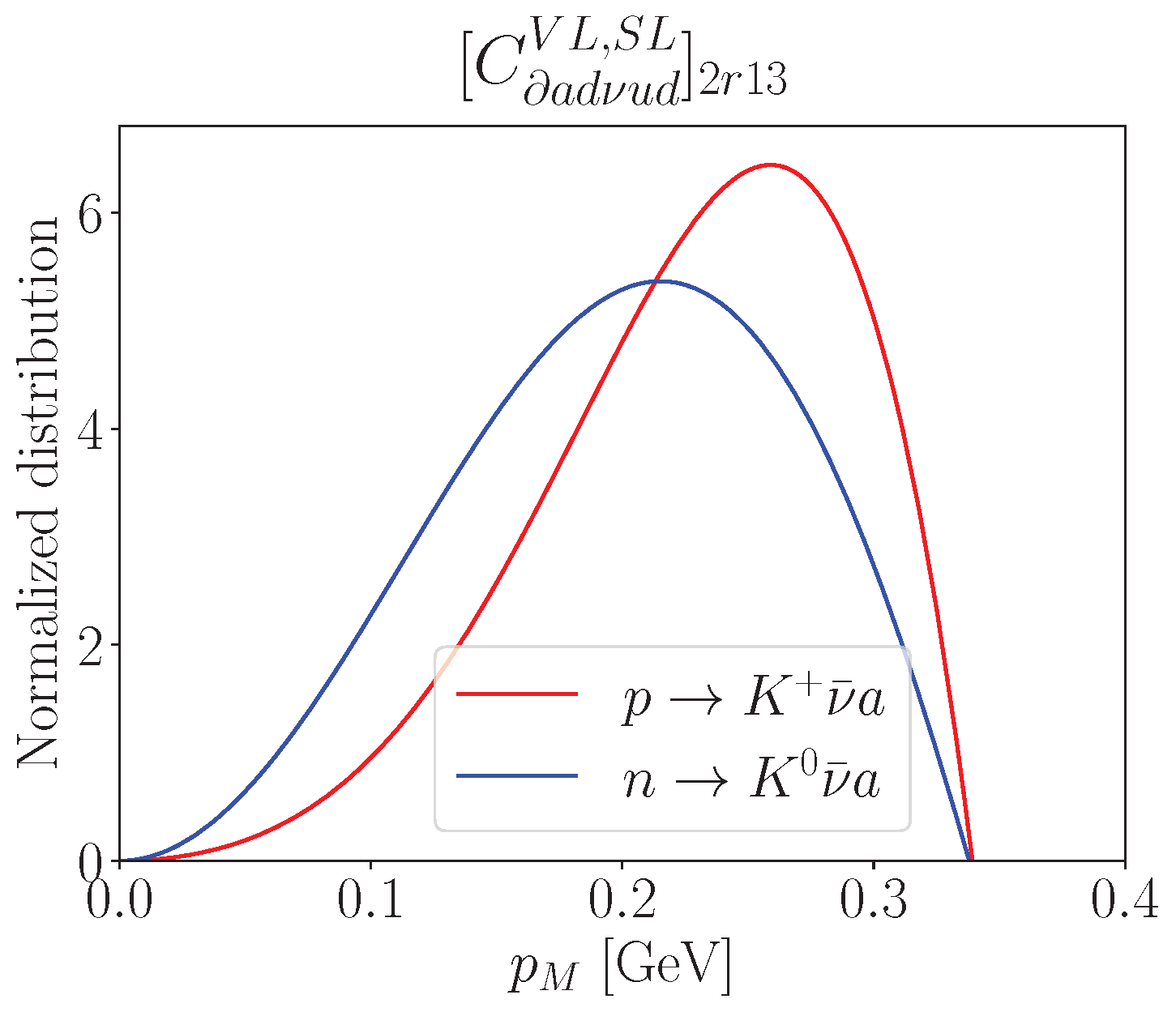}\\
\minigraph{4.5cm}{-0.05in}{}{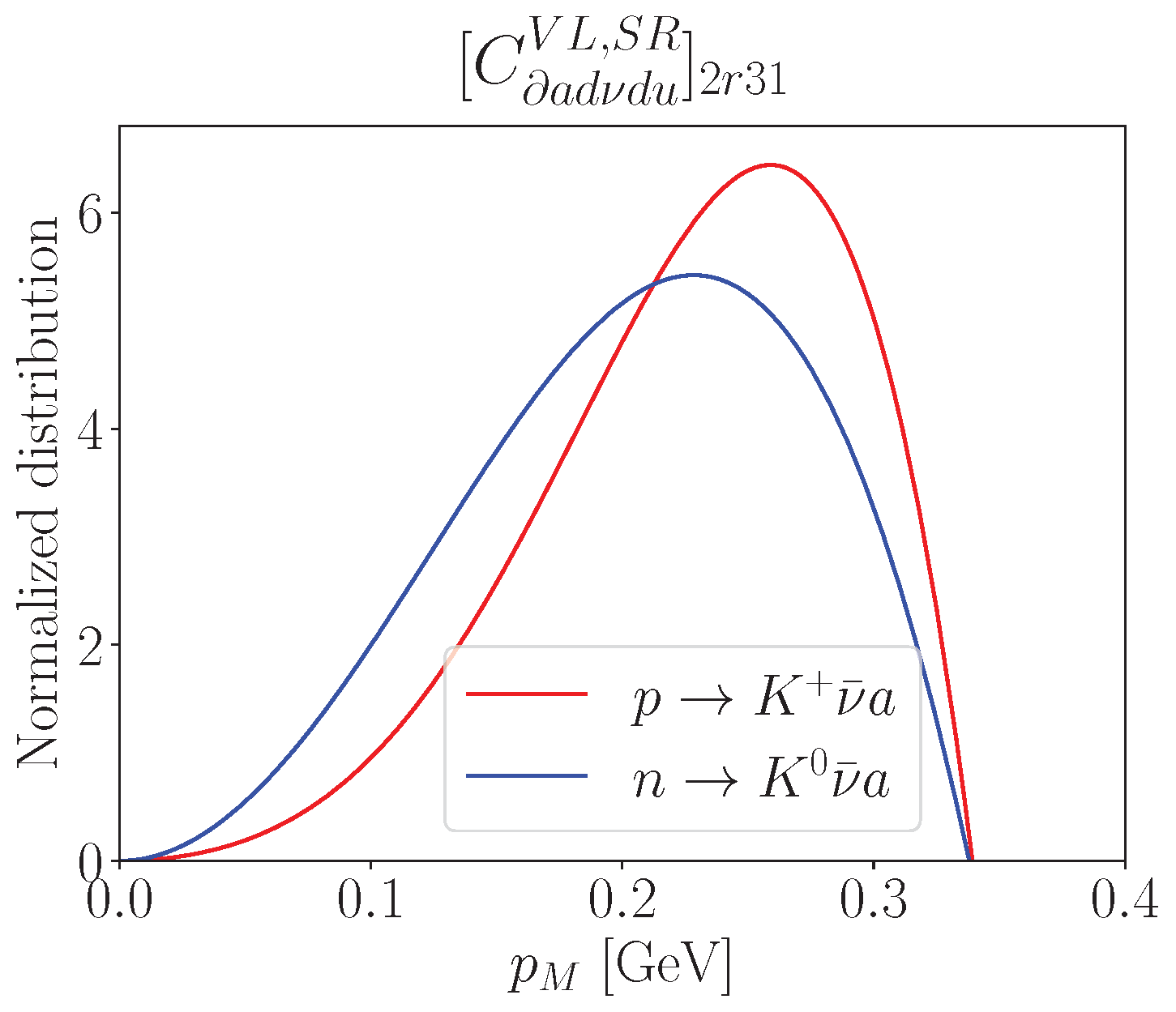}
\minigraph{4.5cm}{-0.05in}{}{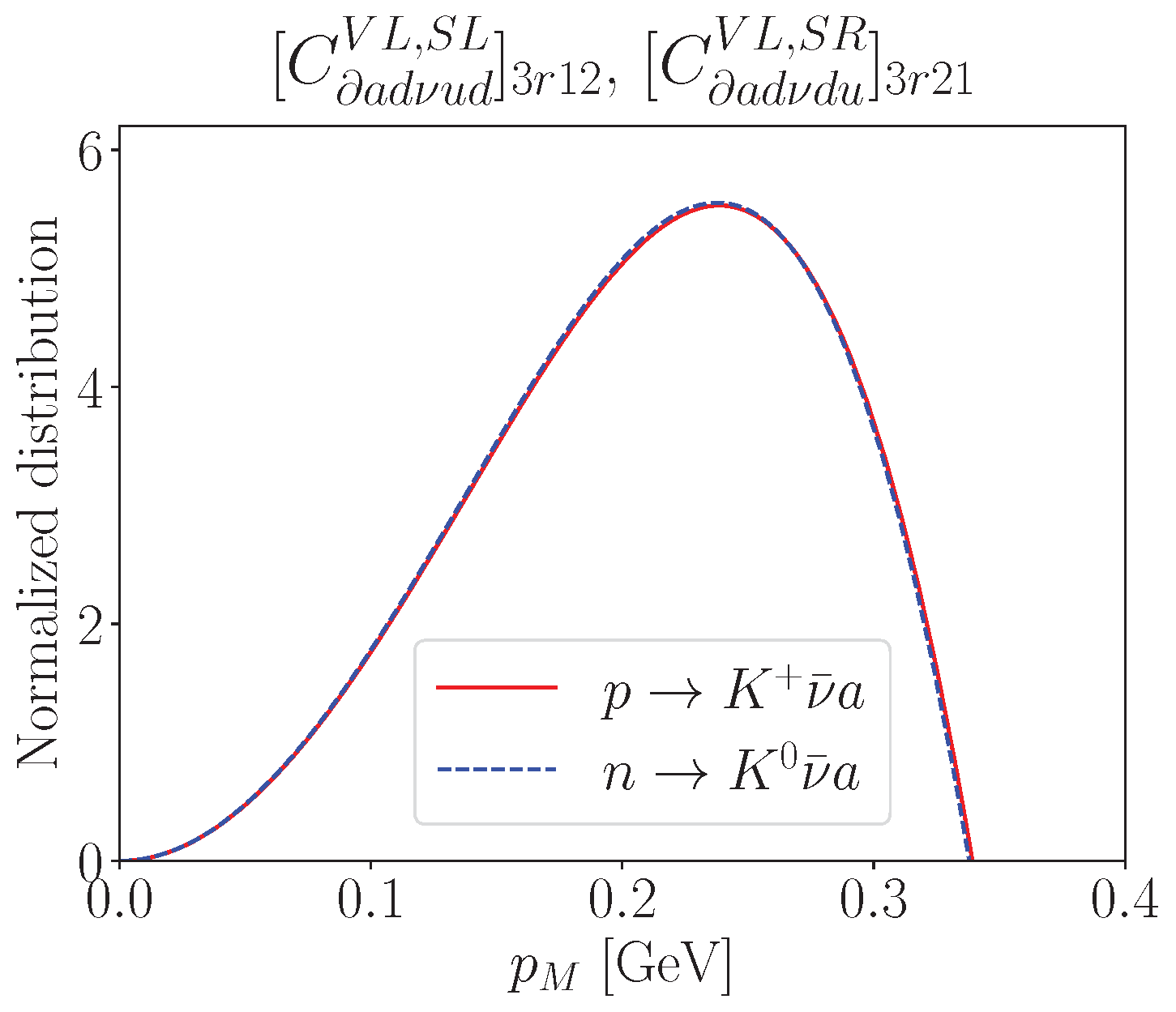}
\minigraph{4.5cm}{-0.05in}{}{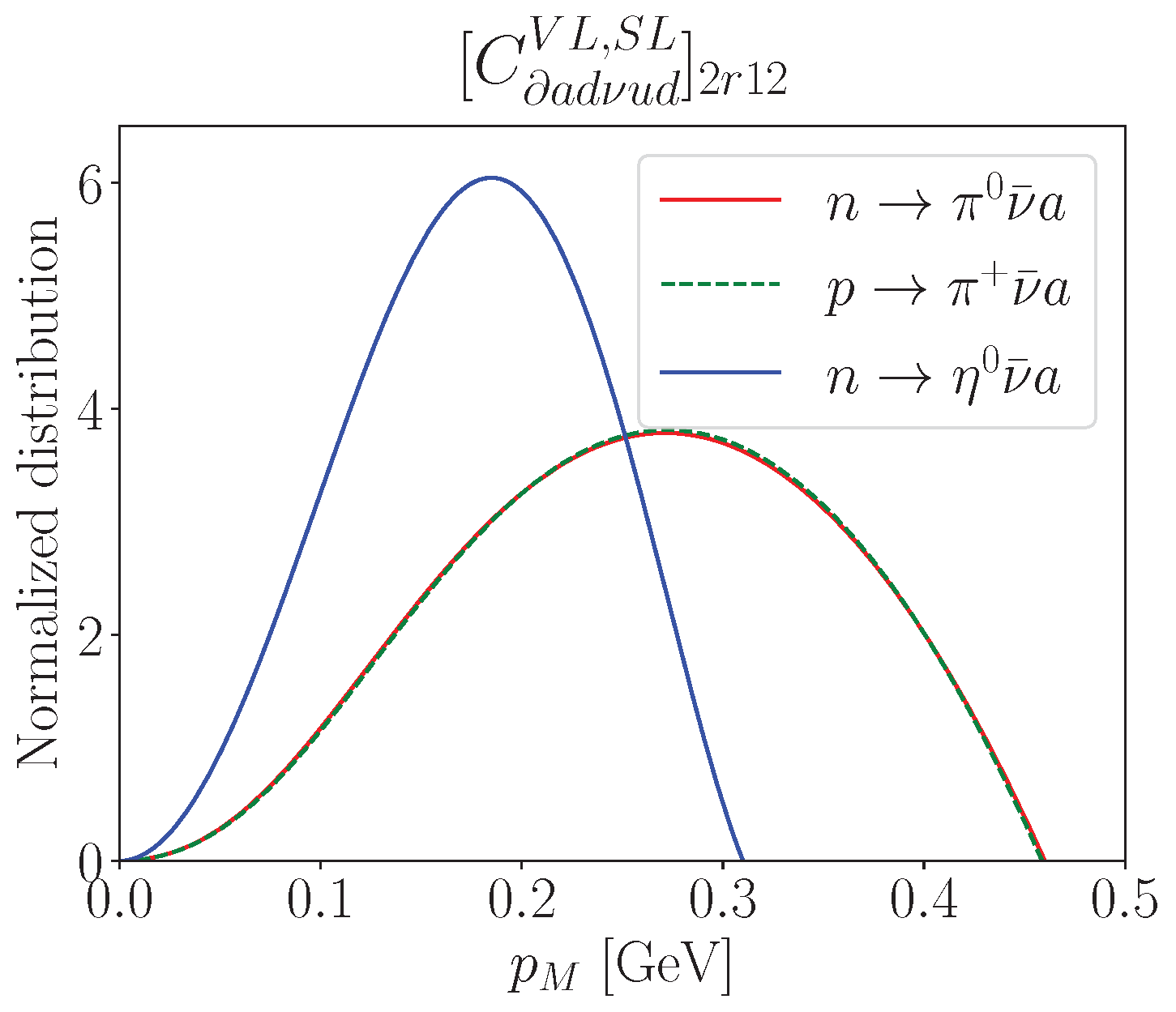}\\
\minigraph{4.5cm}{-0.05in}{}{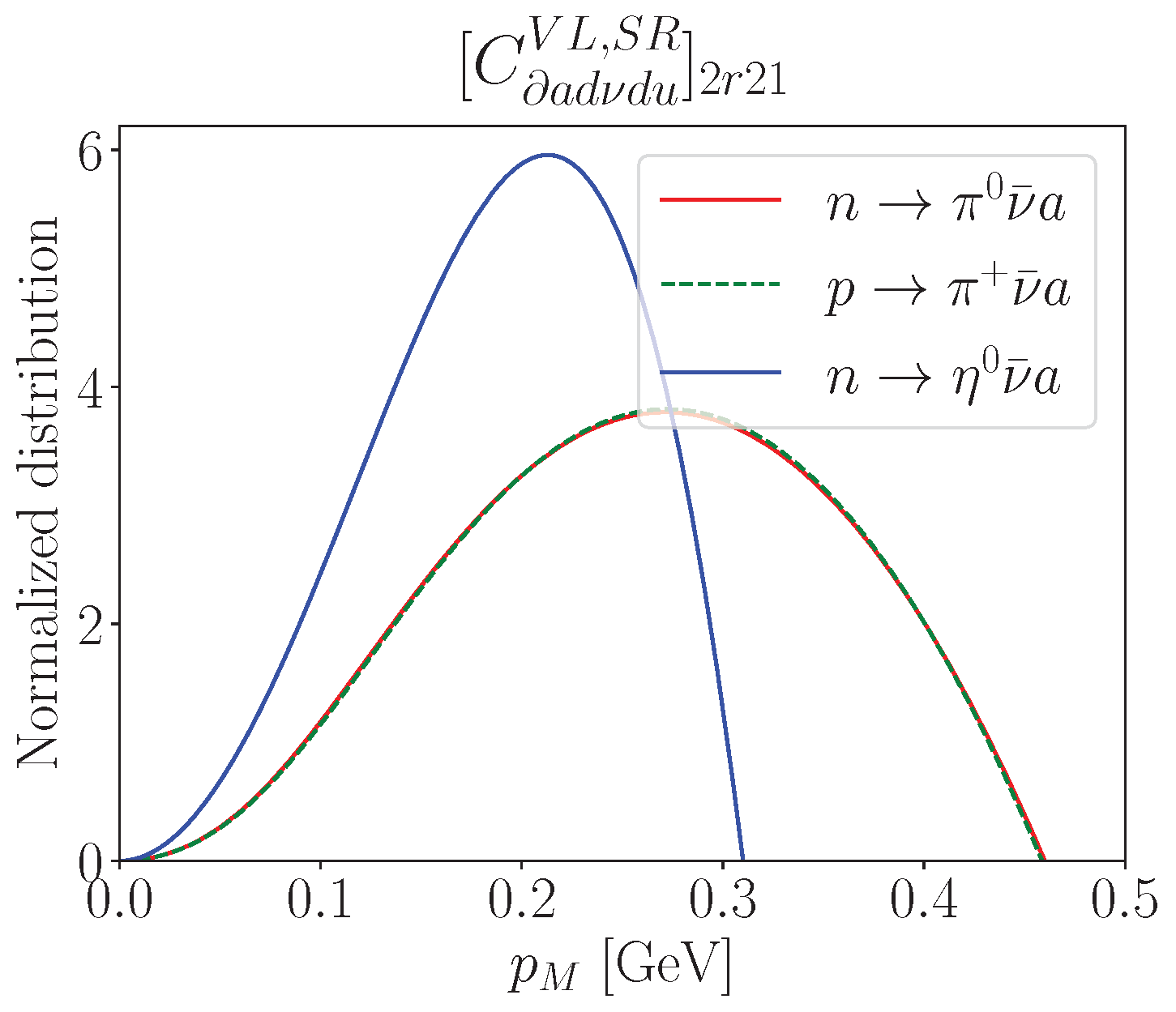}
\minigraph{4.5cm}{-0.05in}{}{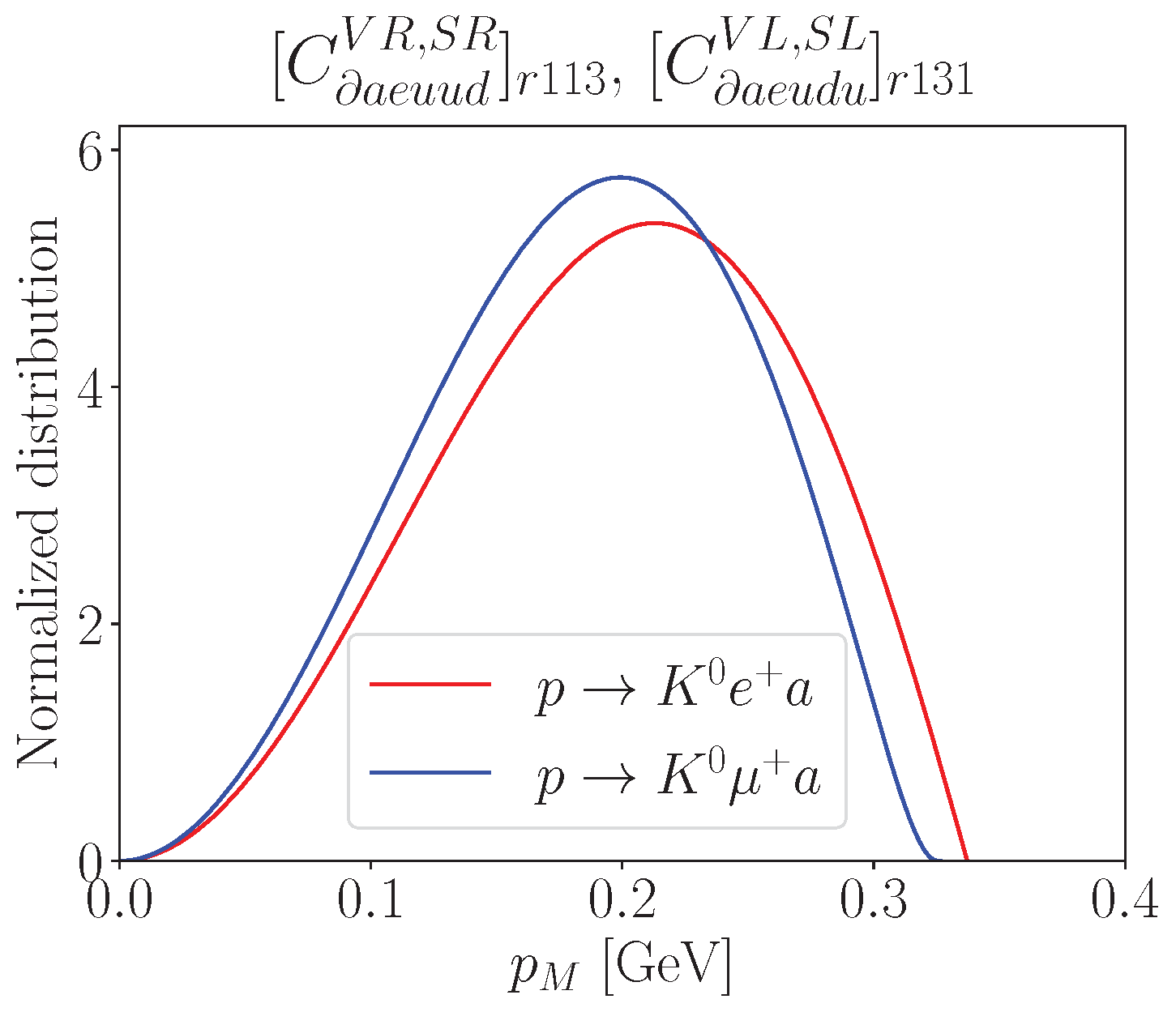}
\minigraph{4.5cm}{-0.05in}{}{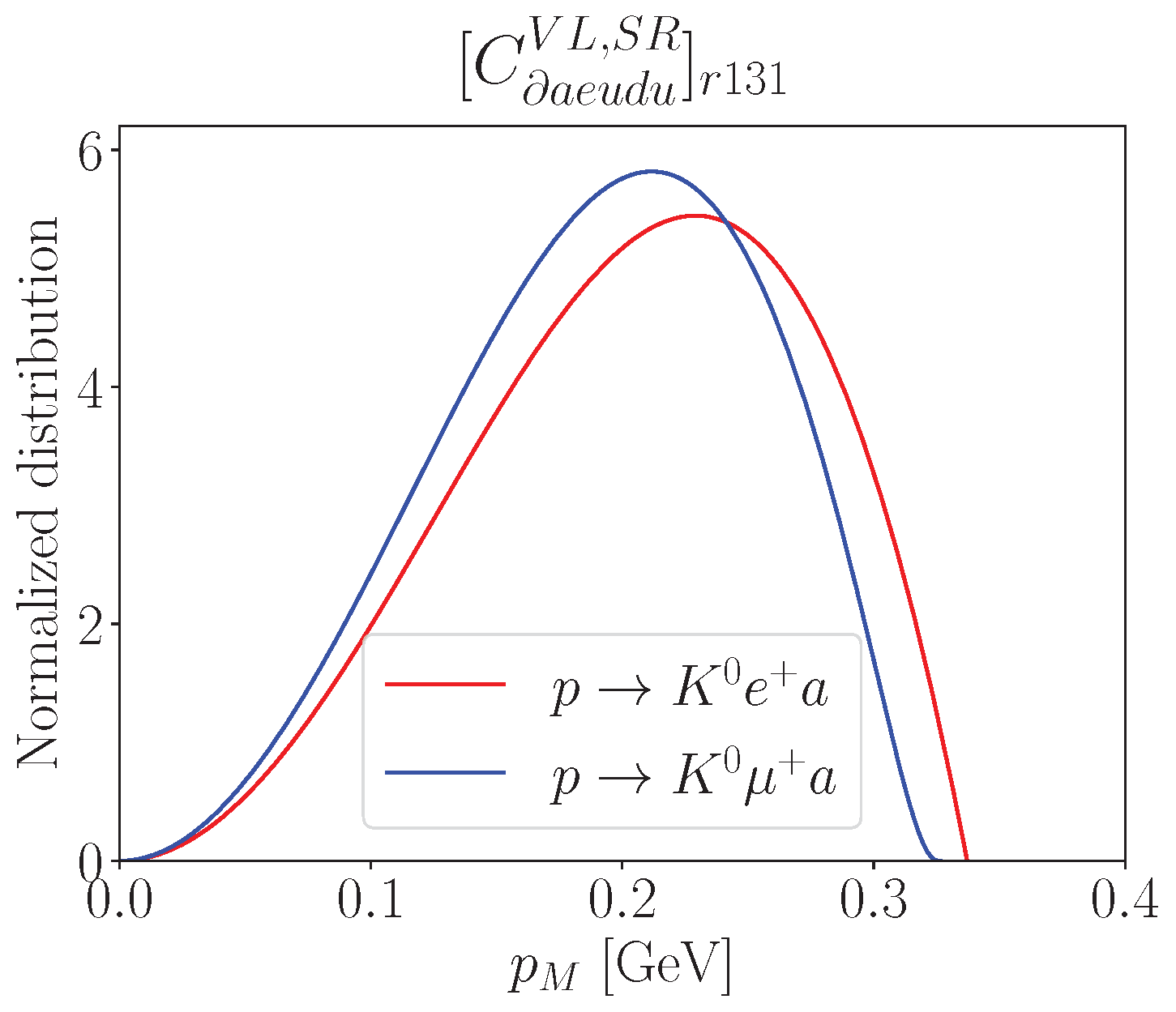}
\caption{The normalized momentum distribution of the final state meson in the rest frame of the nucleon for the processes $N\to M \bar{\ell} a$ with $|\Delta(B-L)|=0$, as labeled in Fig.~\ref{fig:PM_BL2}.
}
\label{fig:PM_BL0}
\end{figure}

The three-body decays $N\to M\ell a$ exhibit different kinematics from the conventional nucleon decays $N\to M\ell$. In Figs.~\ref{fig:PM_BL2} and \ref{fig:PM_BL0}, we show the normalized partial widths of the processes with $|\Delta(B-L)|=2$ and $|\Delta(B-L)|=0$ as a function of the final state meson momentum $p_M\equiv |\vec{p}_M|$ in the rest frame of the nucleon. For each process, we only switch on one WC at a time. The distributions are grouped according to the WC responsible for the processes, as indicated in the plot title. For the plots labeled with multiple WCs, the corresponding coefficients yield the same distribution. For the conventional two-body decays, however, the meson momentum is fixed to be the maximal $p_M$ in the corresponding three-body decays as we set $m_a=0$. This distinct kinematic observable plays as an important distinguishable feature in the presence of ALP.

The current neutrino experiments are under the assumption that the missing energy comes from SM neutrinos and are subject to specific kinematic cuts and selection criteria. As we stated, the meson momentum in the conventional two-body decays is fixed to be the maximal value in the distributions of our Figs.~\ref{fig:PM_BL2} and \ref{fig:PM_BL0}. Thus, the meson momentum in our case would fall outside the range chosen by the experiment and the standard search cut on it may not let our events pass largely. We thus suggest future experiments to broaden their kinematic cut range to search for dark particle such as ALP in nucleon decays.

\subsection{Lower limits on the UV scale and predictions}

The null detection of proton decay $p\to e^+(\mu^+)X$ at Super-K~\cite{Super-Kamiokande:2015pys}, the invisible neutron decay search at KamLAND~\cite{KamLAND:2005pen}, and the invisible decay of $\Lambda^0$ baryon at BESIII~\cite{BESIII:2021slv} can place limits on the WCs of dimension-8 aLEFT BNV operators as well as the lower limits on the UV scale. Tab.~\ref{tab:23bodyLimit} shows the lower limits on the energy scale $1/\sqrt[4]{C}$ for dimension-8 aLEFT operators with $|\Delta (B-L)|=2$ (above the double-line) and $|\Delta (B-L)|=0$ (below the double-line). We assume one non-vanishing operator at a time. These bounds lead to the lower limits of some nucleon partial lifetimes or the upper limit of $\Sigma^0$ decay branching fraction induced by the same operator, as shown in the last column of Tab.~\ref{tab:23bodyLimit}. One can see that the invisible decay of $\Lambda^0$ baryon can only constrain the scale as high as $30-40$ GeV. As the lifetime of $\Sigma^0$ is two orders of magnitude smaller than that of $\Lambda^0$, this constraint pushes ${\rm BR}(\Sigma^0\to \nu(\bar{\nu})a)$ lower than $7.6\times 10^{-14}$. On the other hand, the proton decay at Super-K gives a much stronger lower limit on the UV scale $\sim 10^7$ GeV. The invisible decay of neutron also constrains the scale close to $10^7$ GeV. They confine the partial lifetime of the relevant proton or neutron three-body decays longer than $10^{30}-10^{37}$ years.

\begin{table}[htbp!]
\centering
\renewcommand{\arraystretch}{1.2}
\hspace*{-0.05\columnwidth}\resizebox{1.1\columnwidth}{!}{
\begin{tabular}{c|c|c|c}
\hline
Process & aLEFT WC & $1/\sqrt[4]{C}$ & Prediction 
\\
\hline
$n\to \nu a$ & $[C^{VL,SL}_{\partial a \nu ddu}]_{r221}$ & $8.8\times 10^6$ GeV & $\tau(n\to \pi^0\nu a)>1.5\times 10^{31}~{\rm years}$\\
& & & $\tau(n\to \eta^0 \nu a)>4.7\times 10^{32}~{\rm years}$\\
& & & $\tau(p\to \pi^+ \nu a)>7.6\times 10^{30}~{\rm years}$\\
$\Lambda^0\to \nu a$ & $[C^{VL,SL}_{\partial a \nu ddu}]_{r231}$ & 34.5 GeV & ${\rm BR}(\Sigma^0\to \nu a)<7.4\times 10^{-14}$\\
 & $[C^{VL,SL}_{\partial a \nu ddu}]_{r321}$ & 41.0 GeV\\
\hline
\hline
$p\to e^+ a$ & $[C^{VL,SL}_{\partial a eudu}]_{1121}$, $[C^{VR,SR}_{\partial a euud}]_{1112}$, $[C^{VL,SR}_{\partial a eudu}]_{1121}$ & $2.2\times 10^7$ GeV & $\tau(p\to \pi^0 e^+ a)>2.0\times 10^{34}~{\rm years}$\\
& & & $\tau(p\to \eta^0 e^+ a)>1.8\times 10^{37}~{\rm years}$\\
& & & $\tau(n\to \pi^- e^+ a)>9.9\times 10^{33}~{\rm years}$\\
$p\to \mu^+ a$ & $[C^{VL,SL}_{\partial a eudu}]_{2121}$, $[C^{VR,SR}_{\partial a euud}]_{2112}$, $[C^{VL,SR}_{\partial a eudu}]_{2121}$ & $2.0\times 10^7$ GeV & $\tau(p\to \pi^0 \mu^+ a)>1.2\times 10^{34}~{\rm years}$\\
& & & $\tau(p\to \eta^0 \mu^+ a)>1.4\times 10^{37}~{\rm years}$\\
& & & $\tau(n\to \pi^- \mu^+ a)>6.2\times 10^{33}~{\rm years}$\\
$n\to \bar{\nu} a$ & $[C^{VL,SL}_{\partial a d\nu ud}]_{2r12}$, $[C^{VL,SR}_{\partial a d\nu du}]_{2r21}$ & $8.8\times 10^6$ GeV & $\tau(n\to \pi^0\bar{\nu} a)>1.5\times 10^{31}~{\rm years}$\\
& & & $\tau(n\to \eta^0 \bar{\nu} a)>1.3\times 10^{34}~{\rm years}$\\
& & & $\tau(p\to \pi^+ \bar{\nu} a)>7.6\times 10^{30}~{\rm years}$\\
$\Lambda^0\to \bar{\nu}a$ & $[C^{VL,SL}_{\partial a d\nu ud}]_{2r13}$, $[C^{VL,SR}_{\partial a d\nu du}]_{2r31}$ & 34.5 GeV & ${\rm BR}(\Sigma^0\to \bar{\nu} a)<7.4\times 10^{-14}$\\
 & $[C^{VL,SL}_{\partial a d\nu ud}]_{3r12}$, $[C^{VL,SR}_{\partial a d\nu du}]_{3r21}$ & 41.0 GeV\\
\hline
\end{tabular}
}
\caption{
Lower limits on the energy scale $1/\sqrt[4]{C}$ for dimension-8 aLEFT operators with $|\Delta (B-L)|=2$ (above the double-line) and $|\Delta (B-L)|=0$ (below the double-line) from $p\to e^+(\mu^+)X$, $n\to {\rm invisible}$ or $\Lambda^0\to {\rm invisible}$. We assume that one operator does not vanish at a time. We translate the lower limits on the energy scale into lower limits on the lifetime of three-body nucleon decays and an upper limit on the branching ratio of invisible baryon-number-violating $\Sigma^0$ decays which are shown in the last column. }
\label{tab:23bodyLimit}
\end{table}

\section{Conclusion}
\label{sec:Con}

The search for baryon number violation in nucleon decay is an intriguing probe of NP beyond the SM. The future neutrino experiments bring an increased sensitivity to many BNV nucleon decay modes.
The BNV nucleon decays also offer distinct probe of dark sector states, such as axion or ALP. The invisible ALP induces distinguishing signals and kinematics from the conventional BNV nucleon decays.

In this work, we investigate the ALP EFTs with baryon number violation and the impact of light ALP on BNV nucleon decays. We revisit the dimension-8 BNV operators in the extended SMEFT and LEFT with an ALP field respecting shift symmetry. The aLEFT operators with $|\Delta (B-L)|=2$ and $|\Delta (B-L)|=0$ are matched to the BChPT. We obtain the effective chiral Lagrangian and the BNV interactions between ALP and baryons/mesons. The nucleon decay rates are then calculated. Our main results are summarized as follows.
\begin{itemize}
\item We correct the BNV operator basis of aLEFT in past literature and provide a UV model example.
\item The ALP interactions in BChPT lead to two-body baryon decays $B\to \ell~({\rm or}~\nu)~a$ and three-body nucleon decays $N\to M~\ell~({\rm or}~\nu)~a$.
\item The search for invisible decay of $\Lambda^0$ baryon at BESIII weakly constrains the scale of relevant aLEFT operators with strange quark to $30-40$ GeV.
\item The proton decay search of $p\to e^+(\mu^+)X$ at Super-K places a strong lower limit on the UV scale as $10^7$ GeV. The search for the invisible decay of the neutron at KamLAND also constrains the scale close to $10^7$ GeV.
\item We show the predictions of some other baryon/nucleon decays induced by the same operator. The distinct distributions of kinematic observable in the three-body decays are also presented.
\end{itemize}

\acknowledgments
We would like to thank Guilherme Guedes, Jonathan Kley, Jasper Roosmale Nepveu, Hao-Lin Wang and Jianghao Yu for useful discussions. T.~L. is supported by the National Natural Science Foundation of China (Grant No. 12375096, 12035008, 11975129) and ``the Fundamental Research Funds for the Central Universities'', Nankai University (Grants No. 63196013). M.~S. acknowledges support by the Australian Research Council Discovery Project DP200101470. C.~Y.~Y. is supported in part by the Grants No. NSFC-11975130, No. NSFC-12035008, No. NSFC-12047533, the Helmholtz-OCPC International Postdoctoral Exchange Fellowship Program, the National Key Research and Development Program of China under Grant No. 2017YFA0402200, the China Postdoctoral Science Foundation under Grant No. 2018M641621, and the Deutsche Forschungsgemeinschaft (DFG, German Research Foundation) under Germany's Excellence Strategy --- EXC 2121 ``Quantum Universe'' --- 390833306.

\appendix


\section{The transformation properties of the BNV aLEFT operators}
\label{app:trans}

We decompose the tensor products in the aLEFT operators of interest and derive the irreducible representations under the quark flavor group $SU(3)_L\times SU(3)_R$.
The $|\Delta (B-L)|=2$ operators can be rewritten in terms of the quark multiplet $q=(u,d,s)$ as follows
\begin{align}
\mathcal{O}^{VL,SL}_{\partial a \nu ddu} & \to \partial_\mu a \epsilon^{\alpha\beta\gamma}
(\bar \nu_L \gamma^\mu q_{L,\alpha i})(\bar q_{L,\beta k}^c q_{L,\gamma 1}) \epsilon^{ik1}
+ \partial_\mu a \epsilon^{\alpha\beta\gamma} (\bar \nu_L \gamma^\mu q_{L,\alpha l})(\bar q_{L,\beta k}^c q_{L,\gamma 1}) (\epsilon^{jk1}\delta^l_i -\frac13 \epsilon^{lk1} \delta^j_i)\nonumber \\
&\sim(1+8,1)\;,\\
\mathcal{O}^{VL,SL}_{\partial a eddd}& \to
\partial_\mu a \epsilon^{\alpha\beta\gamma} (\bar e_L \gamma^\mu q_{L,\alpha l} )(\bar q_{L,\beta j}^c q_{L,\gamma k}) \epsilon^{mjk}\delta^{l}_i \sim(8,1) \;,\\
\mathcal{O}^{VR,SR}_{\partial a e ddd} & \to \partial_\mu a \epsilon^{\alpha\beta\gamma} (\bar e_R \gamma^\mu q_{R,\alpha l }) (\bar q^c_{R,\beta j} q_{R,\gamma k})\epsilon^{mjk}\delta^{l}_i \sim(1,8)\;,
\end{align}
where the flavor index $1$ denotes the up quark and $(i,j,k,l,m) \in \{2,3\}$ the light down-type quarks $d,s$.
The $|\Delta (B-L)|=0$ operators become
\begin{align}
\mathcal{O}^{VR,SR}_{\partial a euud}& \to
\partial_\mu a \epsilon^{\alpha\beta\gamma} (\bar e_L^c \gamma^\mu q_{R,\alpha 1 }) (\bar q^c_{R,\beta 1} q_{R,\gamma k})\epsilon^{m1k}\delta^{1}_i \sim(1,8) \;,\\
\mathcal{O}^{VL,SR}_{\partial a eudu} & \to
\partial_\mu a \epsilon^{\alpha\beta\gamma} (\bar e_R^c \gamma^\mu q_{L,\alpha 1 }) (\bar q^c_{R,\beta k} q_{R,\gamma 1})\epsilon^{mk1}\delta^1_i \sim(3,\bar 3)\;,\\
\mathcal{O}^{VL,SL}_{\partial a eudu}& \to
\partial_\mu a \epsilon^{\alpha\beta\gamma} (\bar e_R^c \gamma^\mu q_{L,\alpha1} ) (\bar q^c_{L,\beta k} q_{L,\gamma 1})\epsilon^{mk1}\delta^1_i \sim (8,1)\;,\\
\mathcal{O}^{VL,SL}_{\partial a d\nu ud} & \to
\partial_\mu a \epsilon^{\alpha\beta\gamma} (\bar  q^c_{R,\alpha i} \gamma^\mu \nu_L) (\bar q^c_{L,\beta 1} q_{L,\gamma j})\epsilon^{mj1}\sim(\bar 3,3) \;,\\
\mathcal{O}^{VL,SR}_{\partial a d\nu du} & \to
\partial_\mu a \epsilon^{\alpha\beta\gamma} (\bar q^c_{R,\alpha i} \gamma^\mu \nu_L) (\bar q_{R,\beta j}^c q_{R,\gamma 1})\epsilon^{ij1}
+ \partial_\mu a \epsilon^{\alpha\beta\gamma} (\bar q^c_{R,\alpha l} \gamma^\mu \nu_L) (\bar q_{R,\beta j}^c q_{R,\gamma 1}) ( \epsilon^{mj1}\delta^l_i - \frac13 \epsilon^{lj1} \delta^m_i)\nonumber \\
&\sim (1,1+8)\;.
\end{align}

\section{The spin-averaged squared matrix elements of three-body BNV nucleons decays}
\label{app:3body}

The spin-averaged squared matrix element of $N\to M\ell a$ with $|\Delta(B-L)|=2$ is
\begin{eqnarray}
\overline{|\mathcal{M}|^2}=\overline{|\mathcal{M}|^2_A}+\overline{|\mathcal{M}|^2_B}\;,
\end{eqnarray}
with
\begin{eqnarray}
2\overline{|\mathcal{M}|^2_A}&=&|y^N_{Mr,L}|^2(-m_a^2 m_\ell^2 +m_a^2 m_M^2-m_a^2 m_N^2 -m_\ell^2 m_N^2 + m_\ell^2 u + m_N^2 t -tu)\nonumber \\
&+&y^N_{Mr,L} \sum_{B'}m_{B' r,L}^\ast g_{MB'}^{N\ast}{1\over t-m_{B'}^2}\nonumber \\
&&(-m_a^2 m_\ell^2 m_N^2 + m_a^2 m_\ell^2 t -m_a^2 m_M^2 t +m_\ell^4 m_N^2 -m_\ell^2 m_N^2 t -m_\ell^2 t u + t^2u)\nonumber\\
&+&m_N y^N_{Mr,L} \sum_{B'}m_{B' r,L}^\ast g_{MB'}^{N\ast}{-m_{B'}\over t-m_{B'}^2}\nonumber \\
&&(m_a^2 m_M^2-m_a^2 m_N^2+m_a^2 t - m_\ell^4 -m_\ell^2 m_N^2 +2 m_\ell^2 t +m_\ell^2 u + m_N^2 t -t^2-tu)\nonumber\\
&+&y^{N\ast}_{Mr,L} \sum_{B}m_{Br,L} g_{MB}^{N}{1\over t-m_{B}^2}\nonumber \\
&&(-m_a^2 m_\ell^2 m_N^2 + m_a^2 m_\ell^2 t -m_a^2 m_M^2 t +m_\ell^4 m_N^2 -m_\ell^2 m_N^2 t -m_\ell^2 t u + t^2u)\nonumber\\
&+&\sum_B m_{Br,L}g_{MB}^{N}{-1\over t-m_{B}^2} \sum_{B'}m_{B' r,L}^\ast g_{MB'}^{N\ast}{1\over t-m_{B'}^2}\nonumber\\
&&(m_a^2(m_\ell^2((m_N^2-t)^2-m_M^2 m_N^2)-m_M^2 t^2)+(m_\ell^2-t)(m_\ell^2 m_N^2(m_M^2-m_N^2+t)\nonumber\\
&&-t(m_M^2 m_N^2-m_N^2 u +tu)))\nonumber\\
&+&\sum_B m_{Br,L}g_{MB}^{N}{-m_B\over t-m_{B}^2} \sum_{B'}m_{B' r,L}^\ast g_{MB'}^{N\ast}{m_{B'}\over t-m_{B'}^2}\nonumber\\
&&(m_a^2(-m_\ell^2 m_M^2-m_M^2 m_N^2+(m_N^2-t)^2)+(m_\ell^2-t)(m_\ell^2(m_M^2+m_N^2-t)\nonumber\\
&&+t(-m_M^2+t+u)+m_N^4-m_N^2(2t+u)))\nonumber\\
&+&m_N y^{N\ast}_{Mr,L} \sum_{B}m_{Br,L} g_{MB}^{N}{-m_{B}\over t-m_{B}^2}\nonumber \\
&&(m_a^2 m_M^2-m_a^2 m_N^2+m_a^2 t - m_\ell^4 -m_\ell^2 m_N^2 +2 m_\ell^2 t +m_\ell^2 u + m_N^2 t -t^2-tu)\nonumber\\
&+&m_N(\sum_B m_{Br,L}g_{MB}^{N}{1\over t-m_{B}^2} \sum_{B'}m_{B' r,L}^\ast g_{MB'}^{N\ast}{m_{B'}\over t-m_{B'}^2}\nonumber\\
&&+\sum_B m_{Br,L}g_{MB}^{N}{m_B\over t-m_{B}^2} \sum_{B'}m_{B' r,L}^\ast g_{MB'}^{N\ast}{1\over t-m_{B'}^2})m_M^2 (m_a^2(m_\ell^2+t)-(m_\ell^2-t)^2)\nonumber\\
&+&(L\to R)_1-(L\to R)_2-(L\to R)_3-(L\to R)_4+(L\to R)_5+(L\to R)_6\nonumber \\
&-&(L\to R)_7+(L\to R)_8\;,
\end{eqnarray}
where $t=(p_a+p_\ell)^2$, $u=(p_\ell+p_M)^2$, $()_i$ means the ``$i$''th term in $2\overline{|\mathcal{M}|^2_A}$, and
\begin{eqnarray}
2\overline{|\mathcal{M}|^2_B}&=&m_\ell y_{Mr,L}^N \sum_{B'}m_{B' r,R}^\ast g_{MB'}^{N\ast}{m_{B'}\over t-m_{B'}^2}m_a^2(m_M^2+m_N^2-t)\nonumber \\
&+&m_\ell y_{Mr,L}^N m_N y_{Mr,R}^{N\ast} 2m_a^2\nonumber\\
&+&m_\ell y_{Mr,L}^N m_N \sum_{B'}m_{B' r,R}^\ast g_{MB'}^{N\ast} {1\over t-m_{B'}^2} m_a^2(m_M^2-m_N^2+t)\nonumber\\
&+&m_\ell y_{Mr,R}^{N\ast} \sum_B m_{Br,L}g_{MB}^{N}{-m_B\over t-m_{B}^2}m_a^2(m_M^2+m_N^2-t)\nonumber\\
&+&m_\ell (\sum_B m_{Br,L}g_{MB}^{N}{1\over t-m_B^2}\sum_{B'}m_{B' r,R}^\ast g_{MB'}^{N\ast}{-m_{B'}\over t-m_{B'}^2}\nonumber\\
&&+\sum_B m_{Br,L}g_{MB}^{N}{-m_B\over t-m_B^2}\sum_{B'}m_{B' r,R}^\ast g_{MB'}^{N\ast}{1\over t-m_{B'}^2})m_a^2(m_M^2(m_N^2+t)-(m_N^2-t)^2)\nonumber\\
&+&m_\ell m_N y_{Mr,R}^{N\ast} \sum_B m_{Br,L}g_{MB}^{N}{-1\over t-m_B^2}m_a^2(m_M^2-m_N^2+t)\nonumber\\
&+&m_\ell m_N (\sum_B m_{Br,L}g_{MB}^{N}{1\over t-m_B^2}\sum_{B'}m_{B' r,R}^\ast g_{MB'}^{N\ast}{-1\over t-m_{B'}^2} t \nonumber\\
&&+ \sum_B m_{Br,L}g_{MB}^{N}{m_B\over t-m_B^2}\sum_{B'}m_{B' r,R}^\ast g_{MB'}^{N\ast}{-m_{B'}\over t-m_{B'}^2}) 2m_a^2 m_M^2\nonumber \\
&-&(L\leftrightarrow R)_1+(L\leftrightarrow R)_2-(L\leftrightarrow R)_3-(L\leftrightarrow R)_4+(L\leftrightarrow R)_5\nonumber\\
&-&(L\leftrightarrow R)_6 +(L\leftrightarrow R)_7\;,
\end{eqnarray}
where $()_i$ means the ``$i$''th term in $2\overline{|\mathcal{M}|^2_B}$.
In the limit of massless ALP, $\overline{|\mathcal{M}|^2_B}$ vanishes.

The spin-averaged squared matrix element of $N\to M\bar{\ell} a$ for $|\Delta(B-L)|=0$ is
\begin{eqnarray}
\overline{|\mathcal{M}|^2}=\overline{|\mathcal{M}|^2_A}+\overline{|\mathcal{M}|^2_B}\;,
\end{eqnarray}
with
\begin{eqnarray}
2\overline{|\mathcal{M}|^2_A}&=& ()_1-()_2-()_3-()_4+()_5+()_6-()_7+()_8\nonumber\\
&+&(L\to R)_1+(L\to R)_2+(L\to R)_3+(L\to R)_4+(L\to R)_5+(L\to R)_6\nonumber \\
&&+(L\to R)_7+(L\to R)_8\;,
\end{eqnarray}
where $()_i$ means the ``$i$''th term in $2\overline{|\mathcal{M}|^2_A}$ of $|\Delta(B-L)|=2$, and
\begin{eqnarray}
2\overline{|\mathcal{M}|^2_B}&=& -()_1+()_2-()_3-()_4+()_5-()_6+()_7\nonumber\\
&+&(L\leftrightarrow R)_1+(L\leftrightarrow R)_2+(L\leftrightarrow R)_3+(L\leftrightarrow R)_4+(L\leftrightarrow R)_5+(L\leftrightarrow R)_6\nonumber \\
&+&(L\leftrightarrow R)_7\;,
\end{eqnarray}
where $()_i$ means the ``$i$''th term in $2\overline{|\mathcal{M}|^2_B}$ of $|\Delta(B-L)|=2$.

The decay width becomes
\begin{eqnarray}
\Gamma(N\to M \ell_r a)&=&{1\over (2\pi)^3}{1\over 32 m_N^3}\int dt \int du\overline{|\mathcal{M}|^2}\;,
\end{eqnarray}
where
\begin{eqnarray}
u_{\rm max}&=&(E_\ell^\ast + E_M^\ast)^2-(\sqrt{E_\ell^{\ast 2}-m_\ell^2}-\sqrt{E_M^{\ast 2}-m_M^2})^2\;,\\
u_{\rm min}&=&(E_\ell^\ast + E_M^\ast)^2-(\sqrt{E_\ell^{\ast 2}-m_\ell^2}+\sqrt{E_M^{\ast 2}-m_M^2})^2\;,\\
E_\ell^\ast &=& (t-m_a^2+m_\ell^2)/(2\sqrt{t})\;,\\
E_M^\ast &=& (m_N^2-t-m_M^2)/(2\sqrt{t})\;,\\
t_{\rm max}&=&(m_N-m_M)^2\;,\\
t_{\rm min}&=&(m_a+m_\ell)^2\;.
\end{eqnarray}

\section{The numerical formulae for BNV baryon decays with $|\Delta (B-L)|=0$}
\label{app:BL0}

\underline{Two-body BNV baryon decays:}

(1) $p\to e^+ (\mu^+) a$
\begin{eqnarray}
\Gamma(p\to e^+ a)
&=&\Big({|[C^{VL,SL}_{\partial a eudu}]_{1121}|^2 + |[C^{VR,SR}_{\partial a euud}]_{1112}|^2+ |[C^{VL,SR}_{\partial a eudu}]_{1121}|^2 \over 1.6\times 10^{-59}~{\rm GeV}^{-8}}\nonumber\\
&&-{{\rm Re}([C^{VL,SL}_{\partial a eudu}]_{1121} [C^{VL,SR}_{\partial a eudu}]_{1121}^\ast) \over 8.0\times 10^{-60}~{\rm GeV}^{-8}} \Big) {1\over 10^{33}~{\rm years}}\;,
\end{eqnarray}

\begin{eqnarray}
\Gamma (p\to \mu^+ a)
&=&\Big({|[C^{VL,SL}_{\partial a eudu}]_{2121}|^2 + |[C^{VR,SR}_{\partial a euud}]_{2112}|^2 + |[C^{VL,SR}_{\partial a eudu}]_{2121}|^2 \over 1.6\times 10^{-59}~{\rm GeV}^{-8}} \nonumber\\
&&-{{\rm Re}([C^{VL,SL}_{\partial a eudu}]_{2121} [C^{VL,SR}_{\partial a eudu}]_{2121}^\ast) \over 8.3\times 10^{-60}~{\rm GeV}^{-8}} \Big) {1\over 10^{33}~{\rm years}}\;,
\end{eqnarray}

(2) $\Sigma^+\to e^+(\mu^+) a$
\begin{eqnarray}
{\rm BR}(\Sigma^+\to e^+ a)&=&3.3\times 10^8~{\rm GeV}^8 (|[C^{VL,SL}_{\partial a eudu}]_{1131}|^2+|[C^{VR,SR}_{\partial a euud}]_{1113}|^2 + |[C^{VL,SR}_{\partial a eudu}]_{1131}|^2 ) \nonumber \\
&&- 6.5\times 10^8~{\rm GeV}^8 {\rm Re}([C^{VL,SL}_{\partial a eudu}]_{1131} [C^{VL,SR}_{\partial a eudu}]_{1131}^\ast)\;,
\end{eqnarray}

\begin{eqnarray}
{\rm BR}(\Sigma^+\to \mu^+ a)&=&3.2\times 10^8~{\rm GeV}^8 (|[C^{VL,SL}_{\partial a eudu}]_{2131}|^2+|[C^{VR,SR}_{\partial a euud}]_{2113}|^2 + |[C^{VL,SR}_{\partial a eudu}]_{2131}|^2 )\nonumber \\
&&- 6.4\times 10^8~{\rm GeV}^8 {\rm Re}([C^{VL,SL}_{\partial a eudu}]_{2131} [C^{VL,SR}_{\partial a eudu}]_{2131}^\ast)\;,
\end{eqnarray}

(3) $n\to \bar{\nu} a$
\begin{eqnarray}
\Gamma (n\to \bar{\nu} a)
&=&\Big({|[C^{VL,SL}_{\partial a d\nu ud}]_{2r12}|^2\over 1.6\times 10^{-59}~{\rm GeV}^{-8}}+{|[C^{VL,SR}_{\partial a d\nu du}]_{2r21}|^2\over 1.6\times 10^{-59}~{\rm GeV}^{-8}}\nonumber \\
&&+{{\rm Re}([C^{VL,SL}_{\partial a d\nu ud}]_{2r12} [C^{VL,SR}_{\partial a d\nu du}]_{2r21}^\ast)\over 7.9\times 10^{-60}~{\rm GeV}^{-8}}\Big){1\over 10^{33}~{\rm years}}\;,
\end{eqnarray}

(4) $\Lambda^0\to \bar{\nu} a$
\begin{eqnarray}
{\rm BR}(\Lambda^0\to \bar{\nu} a)&=&1.5\times 10^8~{\rm GeV}^8 |[C^{VL,SL}_{\partial a d\nu ud}]_{2r13}|^2+5.9\times 10^8~{\rm GeV}^8 {\rm Re}([C^{VL,SL}_{\partial a d\nu ud}]_{2r13} [C^{VL,SL}_{\partial a d\nu ud}]_{3r12}^\ast)\nonumber\\
&&+5.9\times 10^8~{\rm GeV}^8 |[C^{VL,SL}_{\partial a d\nu ud}]_{3r12}|^2+2.9\times 10^8~{\rm GeV}^8 {\rm Re}([C^{VL,SL}_{\partial a d\nu ud}]_{2r13} [C^{VL,SR}_{\partial a d\nu du}]_{2r31}^\ast)\nonumber\\
&&+5.9\times 10^8~{\rm GeV}^8 {\rm Re}([C^{VL,SL}_{\partial a d\nu ud}]_{3r12} [C^{VL,SR}_{\partial a d\nu du}]_{2r31}^\ast)+1.5\times 10^8~{\rm GeV}^8 |[C^{VL,SR}_{\partial a d\nu du}]_{2r31}|^2\nonumber \\
&&+5.9\times 10^8~{\rm GeV}^8 {\rm Re}([C^{VL,SL}_{\partial a d\nu ud}]_{2r13} [C^{VL,SR}_{\partial a d\nu du}]_{3r21}^\ast)\nonumber\\
&&+1.2\times 10^9~{\rm GeV}^8 {\rm Re}([C^{VL,SL}_{\partial a d\nu ud}]_{3r12} [C^{VL,SR}_{\partial a d\nu du}]_{3r21}^\ast)\nonumber \\
&&+5.9\times 10^8~{\rm GeV}^8 {\rm Re}([C^{VL,SR}_{\partial a d\nu du}]_{2r31} [C^{VL,SR}_{\partial a d\nu du}]_{3r21}^\ast)+5.9\times 10^8~{\rm GeV}^8 |[C^{VL,SR}_{\partial a d\nu du}]_{3r21}|^2\;,\nonumber\\
\end{eqnarray}

(5) $\Sigma^0\to \bar{\nu} a$
\begin{eqnarray}
{\rm BR}(\Sigma^0\to \bar{\nu} a) &=& 0.15~{\rm GeV}^8 (|[C^{VL,SL}_{\partial a d\nu ud}]_{2r13}|^2 + |[C^{VL,SR}_{\partial a d\nu du}]_{2r31}|^2) \nonumber \\
&&+0.30~{\rm GeV}^8 {\rm Re}([C^{VL,SL}_{\partial a d\nu ud}]_{2r13} [C^{VL,SR}_{\partial a d\nu du}]_{2r31}^\ast)\;,
\end{eqnarray}

(6) $\Xi^0\to \bar{\nu} a$
\begin{eqnarray}
{\rm BR}(\Xi^0\to \bar{\nu} a) &=& 1.6\times 10^9~{\rm GeV}^8 (|[C^{VL,SL}_{\partial a d\nu ud}]_{3r13}|^2 +  |[C^{VL,SR}_{\partial a d\nu du}]_{3r31}|^2 )\nonumber \\
&&+3.2\times 10^9~{\rm GeV}^8 {\rm Re}([C^{VL,SL}_{\partial a d\nu ud}]_{3r13} [C^{VL,SR}_{\partial a d\nu du}]_{3r31}^\ast)\;,
\end{eqnarray}

\underline{Three-body BNV nucleon decays:}


(1) $p\to \pi^0 e^+(\mu^+) a$

\begin{eqnarray}
\Gamma(p\to \pi^0 e^+ a)
&=&\Big({|[C^{VL,SL}_{\partial a eudu}]_{1121}|^2 + |[C^{VL,SR}_{\partial a eudu}]_{1121}|^2+|[C^{VR,SR}_{\partial a euud}]_{1112}|^2\over 4.0\times 10^{-58}~{\rm GeV}^{-8}}\nonumber\\
&-&{{\rm Re}([C^{VL,SR}_{\partial a eudu}]_{1121}[C^{VL,SL}_{\partial a eudu}]_{1121}^*)\over 2.0\times 10^{-58}~{\rm GeV}^{-8}}\Big){1\over 10^{33}~{\rm years}}\;,
\end{eqnarray}

\begin{eqnarray}
\Gamma(p\to \pi^0 \mu^+ a)
&=&\Big({|[C^{VL,SL}_{\partial a eudu}]_{2121}|^2 + |[C^{VL,SR}_{\partial a eudu}]_{2121}|^2+|[C^{VR,SR}_{\partial a euud}]_{2112}|^2\over 4.6\times 10^{-58}~{\rm GeV}^{-8}}\nonumber\\
&-&{{\rm Re}([C^{VL,SR}_{\partial a eudu}]_{2121}[C^{VL,SL}_{\partial a eudu}]_{2121}^*)\over 2.3\times 10^{-58}~{\rm GeV}^{-8}}\Big){1\over 10^{33}~{\rm years}}\;,
\end{eqnarray}

(2) $p\to \eta^0 e^+(\mu^+) a$

\begin{eqnarray}
\Gamma(p\to \eta^0 e^+ a)
&=&\Big({|[C^{VL,SL}_{\partial a eudu}]_{1121}|^2+|[C^{VR,SR}_{\partial a euud}]_{1112}|^2\over 1.3\times 10^{-56}~{\rm GeV}^{-8}}+{|[C^{VL,SR}_{\partial a eudu}]_{1121}|^2\over 3.6\times 10^{-55}~{\rm GeV}^{-8}}\nonumber\\
&+&{{\rm Re}([C^{VL,SR}_{\partial a eudu}]_{1121}[C^{VL,SL}_{\partial a eudu}]_{1121}^*)\over 3.7\times 10^{-56}~{\rm GeV}^{-8}}\Big){1\over 10^{33}~{\rm years}}\;,
\end{eqnarray}

\begin{eqnarray}
\Gamma(p\to \eta^0 \mu^+ a)
&=&\Big({|[C^{VL,SL}_{\partial a eudu}]_{2121}|^2+|[C^{VR,SR}_{\partial a euud}]_{2112}|^2\over 2.1\times 10^{-56}~{\rm GeV}^{-8}}+{|[C^{VL,SR}_{\partial a eudu}]_{2121}|^2\over 5.5\times 10^{-55}~{\rm GeV}^{-8}}\nonumber\\
&+&{{\rm Re}([C^{VL,SR}_{\partial a eudu}]_{2121}[C^{VL,SL}_{\partial a eudu}]_{1121}^*)\over 5.8\times 10^{-56}~{\rm GeV}^{-8}}\Big){1\over 10^{33}~{\rm years}}\;,
\end{eqnarray}

(3) $p\to K^+ \bar{\nu} a$

\begin{eqnarray}
\Gamma(p\to K^+ \bar{\nu} a)
&=&\Big({|[C^{VL,SL}_{\partial a d\nu ud}]_{2r13}|^2+|[C^{VL,SR}_{\partial a d\nu du}]_{2r31}|^2\over 1.5\times 10^{-55}~{\rm GeV}^{-8}}+{|[C^{VL,SL}_{\partial a d\nu ud}]_{3r12}|^2+|[C^{VL,SR}_{\partial a d\nu du}]_{3r21}|^2\over 6.5\times 10^{-57}~{\rm GeV}^{-8}}\nonumber\\
&+&{{\rm Re}([C^{VL,SL}_{\partial a d\nu ud}]_{2r13} [C^{VL,SL}_{\partial a d\nu ud}]_{3r12}^*)+{\rm Re}([C^{VL,SR}_{\partial a d\nu du}]_{2r31} [C^{VL,SL}_{\partial a d\nu ud}]_{3r12}^*)\over 1.7\times 10^{-56}~{\rm GeV}^{-8}}\nonumber\\
&+&{{\rm Re}([C^{VL,SL}_{\partial a d\nu ud}]_{2r13} [C^{VL,SR}_{\partial a d\nu du}]_{3r21}^*)+{\rm Re}([C^{VL,SR}_{\partial a d\nu du}]_{2r31} [C^{VL,SR}_{\partial a d\nu du}]_{3r21}^*)\over 1.7\times 10^{-56}~{\rm GeV}^{-8}}\nonumber\\
&+&{{\rm Re}([C^{VL,SL}_{\partial a d\nu ud}]_{2r13} [C^{VL,SR}_{\partial a d\nu du}]_{2r31}^*)\over  7.4\times 10^{-56}~{\rm GeV}^{-8}}+{{\rm Re}([C^{VL,SL}_{\partial a d\nu ud}]_{3r12} [C^{VL,SR}_{\partial a d\nu du}]_{3r21}^*)\over 3.2\times 10^{-57}~{\rm GeV}^{-8}}\Big) {1\over 10^{33}~{\rm years}}\;,\nonumber\\
\end{eqnarray}

(4) $n\to \pi^- e^+(\mu^+) a$

\begin{eqnarray}
\Gamma(n\to \pi^- e^+ a)
&=&\Big({|[C^{VL,SL}_{\partial a eudu}]_{1121}|^2+|[C^{VL,SR}_{\partial a eudu}]_{1121}|^2+|[C^{VR,SR}_{\partial a euud}]_{1112}|^2\over 2.0\times 10^{-58}~{\rm GeV}^{-8}}\nonumber\\
&-&{{\rm Re}([C^{VL,SR}_{\partial a eudu}]_{1121}[C^{VL,SL}_{\partial a eudu}]_{1121}^*)\over 1.0\times 10^{-58}~{\rm GeV}^{-8}}\Big){1\over 10^{33}~{\rm years}}\;,
\end{eqnarray}

\begin{eqnarray}
\Gamma(n\to \pi^- \mu^+ a)
&=&\Big({|[C^{VL,SL}_{\partial a eudu}]_{2121}|^2+|[C^{VR,SR}_{\partial a euud}]_{2112}|^2\over 2.3\times 10^{-58}~{\rm GeV}^{-8}}+{|[C^{VL,SR}_{\partial a eudu}]_{2121}|^2\over 2.4\times 10^{-58}~{\rm GeV}^{-8}}\nonumber\\
&-&{{\rm Re}([C^{VL,SR}_{\partial a eudu}]_{2121}[C^{VL,SL}_{\partial a eudu}]_{2121}^*)\over 1.2\times 10^{-58}~{\rm GeV}^{-8}}\Big){1\over 10^{33}~{\rm years}}\;,
\end{eqnarray}

(5) $n\to \pi^0 \bar{\nu} a$

\begin{eqnarray}
\Gamma(n\to \pi^0 \bar{\nu} a)
&=&\Big({|[C^{VL,SL}_{\partial a d\nu ud}]_{2r12}|^2+|[C^{VL,SR}_{\partial a d\nu du}]_{2r21}|^2\over 4.0\times 10^{-58}~{\rm GeV}^{-8}}\nonumber\\
&+&{{\rm Re}([C^{VL,SL}_{\partial a d\nu ud}]_{2r12}[C^{VL,SR}_{\partial a d\nu du}]_{2r21}^*)\over 2.0\times 10^{-58}~{\rm GeV}^{-8}}\Big){1\over 10^{33}~{\rm years}}\;,
\end{eqnarray}

(6) $p\to K^0 e^+(\mu^+) a$

\begin{eqnarray}
\Gamma(p\to K^0 e^+ a)
&=&\Big({|[C^{VL,SL}_{\partial a eudu}]_{1131}|^2+|[C^{VR,SR}_{\partial a euud}]_{1113}|^2\over 1.7\times 10^{-56}~{\rm GeV}^{-8}}+{|[C^{VL,SR}_{\partial a eudu}]_{1131}|^2\over 9.9\times 10^{-57}~{\rm GeV}^{-8}}\nonumber\\
&+&{{\rm Re}([C^{VL,SR}_{\partial a eudu}]_{1131}[C^{VL,SL}_{\partial a eudu}]_{1131}^*)\over 6.5\times 10^{-57}~{\rm GeV}^{-8}}\Big){1\over 10^{33}~{\rm years}}\;,
\end{eqnarray}

\begin{eqnarray}
\Gamma(p\to K^0 \mu^+ a)
&=&\Big({|[C^{VL,SL}_{\partial a eudu}]_{2131}|^2+|[C^{VR,SR}_{\partial a euud}]_{2113}|^2\over 2.4\times 10^{-56}~{\rm GeV}^{-8}}+{|[C^{VL,SR}_{\partial a eudu}]_{2131}|^2\over 1.5\times 10^{-56}~{\rm GeV}^{-8}}\nonumber\\
&+&{{\rm Re}([C^{VL,SR}_{\partial a eudu}]_{2131}[C^{VL,SL}_{\partial a eudu}]_{2131}^*)\over 9.5\times 10^{-57}~{\rm GeV}^{-8}}\Big){1\over 10^{33}~{\rm years}}\;,
\end{eqnarray}

(7) $p\to \pi^+ \bar{\nu} a$

\begin{eqnarray}
\Gamma(p\to \pi^+ \bar{\nu} a)
&=&\Big({|[C^{VL,SL}_{\partial a d\nu ud}]_{2r12}|^2+|[C^{VL,SR}_{\partial a d\nu du}]_{2r21}|^2\over 2.1\times 10^{-58}~{\rm GeV}^{-8}}\nonumber \\
&+&{{\rm Re}([C^{VL,SL}_{\partial a d\nu ud}]_{2r12}[C^{VL,SR}_{\partial a d\nu du}]_{2r21}^*)\over 1.0\times 10^{-58}~{\rm GeV}^{-8}}\Big){1\over 10^{33}~{\rm years}}\;,
\end{eqnarray}

(8) $n\to \eta^0 \bar{\nu} a$

\begin{eqnarray}
\Gamma(n\to \eta^0 \bar{\nu} a)
&=&\Big({|[C^{VL,SL}_{\partial a d\nu ud}]_{2r12}|^2\over 3.5\times 10^{-55}~{\rm GeV}^{-8}}+{|[C^{VL,SR}_{\partial a d\nu du}]_{2r21}|^2\over 1.3\times 10^{-56}~{\rm GeV}^{-8}}\nonumber\\
&-&{{\rm Re}([C^{VL,SL}_{\partial a d\nu ud}]_{2r12}[C^{VL,SR}_{\partial a d\nu du}]_{2r21}^*)\over 3.6\times 10^{-56}~{\rm GeV}^{-8}}\Big){1\over 10^{33}~{\rm years}}\;,
\end{eqnarray}

(9) $n\to K^0 \bar{\nu} a$

\begin{eqnarray}
\Gamma(n\to K^0 \bar{\nu} a)
&=&\Big({|[C^{VL,SL}_{\partial a d\nu ud}]_{2r13}|^2\over 1.6\times 10^{-56}~{\rm GeV}^{-8}}+{|[C^{VL,SR}_{\partial a d\nu du}]_{2r31}|^2\over 1.0\times 10^{-56}~{\rm GeV}^{-8}}+{[C^{VL,SL}_{\partial a d\nu ud}]_{3r12}|^2+|[C^{VL,SR}_{\partial a d\nu du}]_{3r21}|^2\over 6.6\times 10^{-57}~{\rm GeV}^{-8}}\nonumber\\
&-&{{\rm Re}([C^{VL,SL}_{\partial a d\nu ud}]_{2r13} [C^{VL,SL}_{\partial a d\nu ud}]_{3r12}^*)+{\rm Re}([C^{VL,SL}_{\partial a d\nu ud}]_{2r13} [C^{VL,SR}_{\partial a d\nu du}]_{3r21}^*)\over 5.2\times 10^{-57}~{\rm GeV}^{-8}}\nonumber\\
&+&{{\rm Re}([C^{VL,SR}_{\partial a d\nu du}]_{2r31} [C^{VL,SL}_{\partial a d\nu ud}]_{3r12}^*)+{\rm Re}([C^{VL,SR}_{\partial a d\nu du}]_{2r31} [C^{VL,SR}_{\partial a d\nu du}]_{3r21}^*)\over 4.1\times 10^{-57}~{\rm GeV}^{-8}}\nonumber\\
&-&{{\rm Re}([C^{VL,SL}_{\partial a d\nu ud}]_{2r13} [C^{VL,SR}_{\partial a d\nu du}]_{2r31}^*)\over 6.4\times 10^{-57}~{\rm GeV}^{-8}}+{{\rm Re}([C^{VL,SL}_{\partial a d\nu ud}]_{3r12} [C^{VL,SR}_{\partial a d\nu du}]_{3r21}^*)\over 3.3\times 10^{-57}~{\rm GeV}^{-8}}\Big){1\over 10^{33}~{\rm years}}\;.\nonumber\\
\end{eqnarray}

\bibliographystyle{JHEP}
\bibliography{refs}

\end{document}